\documentclass[
    a4paper,12pt,onecolumn,tightenlines, superscriptaddress,
    aps,physrev,nofootinbib,nobalancelastpage,
    amsfonts,amssymb,amsmath
]{revtex4-2}
\usepackage[T1]{fontenc}
\usepackage[utf8]{inputenc}
\usepackage{natbib}
\usepackage{graphicx}
\usepackage{mathtools}
\usepackage{amsfonts} 
\usepackage{amsmath} 
\usepackage{amssymb}
\usepackage{amsthm}
\usepackage{bm}
\usepackage{braket}
\usepackage{color} 
\usepackage{bbm}
\usepackage{hyperref}
\usepackage{subcaption}
\usepackage{import}
\usepackage{microtype}
\usepackage{relsize}
\usepackage{enumitem}
\usepackage{stackengine}
\usepackage[most]{tcolorbox}
\usepackage[dvipsnames]{xcolor}
\graphicspath{{images/}}

\usepackage[retainorgcmds]{IEEEtrantools}

\usepackage{listings}
\definecolor{codebg}{RGB}{40,44,52}
\definecolor{codefg}{RGB}{248,248,242}
\definecolor{keywordcolor}{rgb}{0.7, 0.1, 0.1}   
\definecolor{tacticcolor}{rgb}{0.0, 0.1, 0.6}    
\definecolor{commentcolor}{rgb}{0.4, 0.4, 0.4}   
\definecolor{symbolcolor}{rgb}{0.0, 0.1, 0.6}    
\definecolor{sortcolor}{rgb}{0.1, 0.5, 0.1}      
\definecolor{attributecolor}{rgb}{0.7, 0.1, 0.1} 

\newtcolorbox{codeListingBox}[2][]{
  enhanced,
  breakable,
  boxrule=0.6pt,
  arc=3mm,
  left=4pt,right=4pt,top=4pt,bottom=4pt,
  title={#2},
  fonttitle=\sffamily\small,
  attach boxed title to top left={yshift=-2mm,xshift=4mm},
  boxed title style={sharp corners,boxrule=0pt,boxsep=2pt},
  #1
}

\newcommand{\norm}[1]{\left\lVert#1\right\rVert}
\newcommand{\abs}[1]{|#1|}
\newcommand{\defeq}{\vcentcolon=}

\newcommand{\one}{\mathbbm{1}}
\newcommand{\R}{\mathbb{R}}

\newcommand{\Hilb}{\mathcal{H}}

\DeclareMathOperator{\Tr}{Tr}

\newtheorem{theorem}{Theorem}

\theoremstyle{definition}

\usepackage{dsfont}
\usepackage{thmtools, thm-restate}
\usepackage{physics}
\usepackage{nicefrac}
\usepackage{enumitem}
\usepackage{tcolorbox}
\usepackage{tabularx}

\def \be {\begin{equation}}
\def \ee {\end{equation}}

\def \sofc2{{\cal S}({\mathbb C}^2)}

\def\>{\rangle}
\def\<{\langle}

\newcommand{\mathlib}{\textsc{mathlib}}
\newcommand{\lean}{\textsc{Lean}}
\newcommand{\leanqi}{\textsc{Lean-QuantumInfo}}

\definecolor{darkblue}{rgb}{0.,0.,0.4}

\begin{document}

\title{A Formalization of the Generalized Quantum Stein's Lemma in Lean}

\author{Alex Meiburg}
\affiliation{Perimeter Institute for Theoretical Physics, Waterloo, Ontario N2L 2Y5, Canada}
\affiliation{Institute for Quantum Computing, University of Waterloo, Waterloo, ON, N2L 3G1, Canada}

\author{Leonardo A. Lessa}
\affiliation{Perimeter Institute for Theoretical Physics, Waterloo, Ontario N2L 2Y5, Canada}
\affiliation{Department of Physics and Astronomy, University of Waterloo, Waterloo, Ontario N2L 3G1, Canada}

\author{Rodolfo R. Soldati}
\affiliation{Perimeter Institute for Theoretical Physics, Waterloo, Ontario N2L 2Y5, Canada}
\affiliation{Institute for Quantum Computing, University of Waterloo, Waterloo, ON, N2L 3G1, Canada}
\affiliation{Department of Physics and Astronomy, University of Waterloo, Waterloo, Ontario N2L 3G1, Canada}


\begin{abstract}
The Generalized Quantum Stein's Lemma is a theorem in quantum hypothesis testing that provides an operational meaning to the relative entropy within the context of quantum resource theories. Its original proof was found to have a gap, which led to a search for a corrected proof. We formalize the proof presented in [Hayashi and Yamasaki (2024)] in the Lean interactive theorem prover. This is the most technically demanding theorem in physics with a computer-verified proof to date, building with a variety of intermediate results from topology, analysis, and operator algebra. In the process, we rectified minor imprecisions in [HY24]'s proof that formalization forces us to confront, and refine a more precise definition of quantum resource theory. Formalizing this theorem has ensured that our Lean-QuantumInfo library, which otherwise has begun to encompass a variety of topics from quantum information, includes a robust foundation suitable for a larger collaborative program of formalizing quantum theory more broadly.
\end{abstract}

\maketitle

\section{Introduction}

How does an experimentalist verify the quantum state they have access to in the laboratory? Hypothesis testing is a task in statistics that studies this question. In quantum information theory, this task contrasts the null hypothesis, postulating a state $\rho$, and the alternative hypothesis, postulating a state $\sigma$.

The quantum Stein's lemma is originally a result in hypothesis testing \cite{hiai1991,ogawa2000}, operationally requiring two independent and identically-distributed (i.i.d.) sets of states, copies of $\rho$ and $\sigma$, and determining the asymptotic error rate of mistaking $\rho$ for $\sigma$, when the error of mistaking $\sigma$ for $\rho$ is fixed at some value $\varepsilon > 0$. These two types of errors are termed type-II and type-I, respectively, and the resulting asymptotic rate for the type-II error is the quantum relative entropy $D(\rho\Vert \sigma)$.

A remarkable generalization was attempted in 2010~\cite{brandao2010}, relaxing the i.i.d.\ condition on the alternative hypothesis states $\{\sigma^{\otimes n}\}_n$ to a set of \textit{free states} in a quantum resource theory (QRT), e.g. the set of separable states in the entanglement resource theory~\cite{chitambar2019}. In this generalized scenario, the hypothesis task is to determine the resourcefulness of $\rho$ through binary quantum measurements. 

Besides the importance of the generalized quantum Stein's lemma (GQSL) to hypothesis testing and resource theories, it also carries an interesting story: in 2023, Ref.~\cite{berta2023} found a gap in the original proof of Ref.~\cite{brandao2010}, which subsequently sparked efforts to prove the GQSL, culminating in Refs.~\cite{hayashi2024, lami2025}. These events motivated us to use the lemma as an emblematic target for proof-based quantum information research using \lean.

We formalize this Generalized Quantum Stein's Lemma (GQSL) based on Ref.~\cite{hayashi2024} in the \lean\ Theorem Prover~\cite{moura2021,demoura2015,2025d}.
In order to achieve this goal, the underlying formal structure of quantum information had to be built, which in turn required a more basic and foundational mathematical structure. This sequence of dependencies demonstrates the unique environment for doing proof work in \lean\ (or other theorem-proving languages).

We use the extensive library known as \mathlib~\cite{2025c,themathlibcommunity2020} to cover the basic and foundational mathematics required, and alongside the GQSL we build the \leanqi\ library~\cite{meiburg2025} to support its proof. As of October 2025, the library has over 1000 theorems, 250 definitions, and 15,000 lines of code.

We believe in the long-term goal of formalizing quantum information extensively in \lean, much like the way branches of mathematics are formalized in \mathlib. Likewise, we believe that the proof-based nature of quantum information theory makes it especially amenable to benefit from formalization compared to other subfields of physics.



The organization of the remainder of this manuscript is as follows. We provide background on proof formalization and on the GQSL in Sec.~\ref{sec:background}. We illustrate the former by discussing a formalized proof of the no-cloning theorem in Sec.~\ref{sec:no-cloning}. The outcomes of our formalization are laid out in Sec.~\ref{sec:outcomes_of_the_formalization}, where we formulate the main theorem in \lean, and comment on how it can be interpreted and fits into the larger context of \mathlib\ and of the original proof. We also mention what aspects of the formalization remain. Sec.~\ref{sec:design_and_development_choices} elaborates on details of the Quantum Resource Theories considered, technical choices that had to be made to facilitate, improve or enable the work, and intricacies that the process of formalization makes evident and could otherwise pass unnoticed. 


\section{Background}\label{sec:background}

\subsection{Proof formalization}
\begin{figure}
    \centering
    \includegraphics[width=0.6\linewidth]{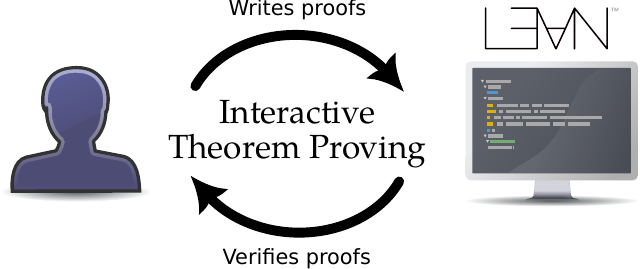}
    \caption{An interacting theorem proving system, such as \lean, is a tool with which the user constructs a proof, and the computer formally verifies it.
    }
    \label{fig:ITP}
\end{figure}

Interactive theorem proving, also known as {\em automated} theorem proving, is the use of computers to construct or check formal proofs of mathematical statements \cite{harrison_handbook_2009}. The human user interacts with the computer by providing coding instructions on the steps of a proof, while receiving feedback in the form of, for example, the correctness and the current state of the proof (See Fig. \ref{fig:ITP}). In contrast to symbolic computation or numerical simulation, proof verification systems operate inside a formal logical framework: every theorem is derived from axioms and inference rules, ensuring that correctness is guaranteed by construction.

\lean\ is one of the leading interactive theorem provers. It is based on the calculus of constructions, a powerful type-theoretic foundation that unifies programming and logic. In \lean, mathematical objects, propositions, and proofs are all represented within the same typed language, a perspective often summarized as ``propositions as types''.

Several other interactive theorem provers exist, such as Rocq (formerly Coq), Isabelle/HOL and Agda, each with distinct logical foundations and proof styles. We chose \lean\ in part because of \mathlib, an extensive community-driven library of formalized mathematics . \mathlib\ covers a wide range of fields, with linear algebra, analysis and topology being the most useful for this project. In addition, \lean\ provides a toolbox of proof tactics that automates many proof steps, leaving only the more non-trivial ones for the user to write. For example, one can prove $1 + 1 = 2$ either by invoking the \lstinline{Nat.succ_eq_add_one} theorem to replace $1 + 1$ with the successor of $1$ --- which is equal to $2$ by definition --- or by simply applying the reflexivity tactic \lstinline{rfl}.


We emphasize that \lean\ is entirely different from a modern large language model, which can produce plausible but unreliable arguments. On the contrary, every proof accepted by \lean\ is verified down to the most basic axioms of mathematics by a trusted kernel. In fact, several mathematical formalization projects have been completed using \lean\ and \mathlib~\cite{lean_papers}, with many others underway.

It is informative to go through steps of a simple yet sufficiently non-trivial proof. One such statement, with broad and important consequences to quantum information science, is the no-cloning theorem \cite{wootters1982}. In its simplest form, the proof of no-cloning is straightforward and results from basic facts of quantum theory. To illustrate the work process of formalization in \lean, we first reproduce a textbook proof in natural language, and then compare it with the corresponding derivation in \lean. In Appendix \ref{app:other_examples}, we provide further examples of theorems relevant to quantum information theory that are proved in the \leanqi\ library

\subsubsection{Example: No-cloning theorem}\label{sec:no-cloning}
Let $\mathcal{H}$ be a $d$-dimensional Hilbert space, and consider two state vectors $\ket{\psi}$, and $\ket{\phi} \in \mathcal{H}$, that are distinct but otherwise arbitrary. To capture the distinction property of $\ket{\psi}$ and $\ket{\phi}$, we state $\braket{\psi}{\phi} < 1$. Consider further a second instance of $\mathcal{H}$, and an arbitrary fiducial state vector $\ket{f}$.

Let $U$ be a unitary acting on the composite Hilbert space $\mathcal{H}^{\otimes 2} = \mathcal{H} \otimes \mathcal{H}$, and assume it implements some cloning state transition for the given state vectors. That is,
\begin{equation}\label{eq:no-cloning}
\begin{split}
    U \colon{}& \mathcal{H}^{\otimes 2} \to\mathcal{H}^{\otimes 2} , \\
    & U \ket{\psi}\ket{f} = \ket{\psi}\ket{\psi} \\
    & U \ket{\phi}\ket{f} = \ket{\phi}\ket{\phi} ,
\end{split}
\end{equation}
where composite state vectors such as $\ket{\psi}\ket{f}$ are shorthand for $\ket{\psi}\otimes\ket{f} \in \mathcal{H}^{\otimes 2}$. We show how these requirements lead to an equation whose set of solutions highly restricts the states that can be cloned.

Examine the inner product of the two output state vectors above, $(\bra{\psi}\otimes\bra{\psi})(\ket{\phi}\otimes\ket{\phi})$. By construction, we can write this as the following amplitude:
\begin{equation}\label{eq:no-cloning2}
    (\bra{\psi}\otimes\bra{\psi})(\ket{\phi}\otimes\ket{\phi}) = \braket{\psi}{\phi}\braket{\psi}{\phi}  = \braket{\psi}{\phi}^2,
\end{equation}
where on the right-hand side the inner products are the ones defined on each tensor factor $\mathcal{H}$.

Because Eq.~\eqref{eq:no-cloning} holds, we can work backwards and introduce the unitary map on the left-hand side above. This yields the alternative equality:
\begin{equation}
    (\bra{\psi}\otimes\bra{\psi})(\ket{\phi}\otimes\ket{\phi}) =
    (\bra{\psi}\otimes\bra{f}) (U^\dag U \ket{\phi}\otimes\ket{f}) .
\end{equation}
The unitary multiplication simplifies, $U^\dag U = \one$,
and we are left with
\begin{equation}\label{eq:no-cloning3}
    (\bra{\psi}\otimes\bra{\psi})(\ket{\phi}\otimes\ket{\phi}) = \braket{\psi}{\phi}\braket{f}{f} .
\end{equation}
Because $\ket{f}$ is a state vector, it is normalized and so $\braket{f} = 1$. Comparing expressions~\eqref{eq:no-cloning2} and~\eqref{eq:no-cloning3} yields
\begin{equation}\label{eq:no-cloning4}
    \braket{\psi}{\phi} (\braket{\psi}{\phi} - 1) = 0 ,
\end{equation}
which has only $\braket{\psi}{\phi} = 0$ and $\braket{\psi}{\phi} = 1$ as solutions. We originally assumed that the two states are distinct, so $\braket{\psi}{\phi} = 1$ is not permissible. We are left with
\begin{equation}
    \braket{\psi}{\phi} = 0 .
\end{equation}
In summary, by defining the action of the cloning unitary, we arrive at the conclusion that it can only be satisfied if $\ket{\psi}$ and $\ket{\phi}$ are orthogonal states. Hence there is no cloning unitary, universal for all state vectors.

What does the proof look like in \lean? It starts with statements following the keyword \lstinline{theorem} and its name, the collection of variables that will be used in the argument, and assumed truths: these are proofs as terms of their corresponding proposition type (See the code listing at the end of this section). For instance, we have hypothesis \lstinline{hψ}, which is a proof of the equality \lstinline{U ◃ pure (ψ ⊗ f) = pure (ψ ⊗ ψ))}. This represents the assumption that $U$ acting on $\ket{\psi}\ket{f}$ gives $\ket{\psi} \ket{\psi}$. 

The three assumptions --- \lstinline{hψ}, \lstinline{hφ} and \lstinline{H} --- are followed by the statement of the theorem: \lstinline{⟪pure ψ, pure φ⟫ = (0 : ℝ)}, i.e. $\ket{\psi}$ and $\ket{\varphi}$ are orthogonal. In \lean, this is the type of \lstinline{no_cloning}, and it. In the language of dependent-type theory, this type is itself also a term, since it \emph{inhabits} the larger type \lstinline{Prop}, of propositions.

The proof starts after the keyword \lstinline{by}. We are given the goal of constructing a term of the type above, which can be done interactively. In practice, this means that, at all steps of the proof, the human prover knows which assumptions are available to use, and what is the current goal. The rules for manipulating the goal and assumptions mainly involve \emph{tactics} --- built-in program that modify the proof state --- and other theorems available in the context. 

These programs have different levels of automation. Take for example the tactic \lstinline{simp}, which stands for \emph{simplify}. When invoked, this tactic searches for theorems flagged with \lstinline{@[simp]} and attempts to apply them to the current goal. A useful step may not always be found, or may not always yield the best outcome. A more direct tactic is \lstinline{rw}, for \emph{rewrite}, which takes an identity (e.g. \lstinline{A = B} or \lstinline{A ↔ B}) as input and updates the goal with the conclusion of that identity. In the proof of the no-cloning theorem below we use it in \lstinline{rw [mul_eq_zero] at h3}. This line targets the term \lstinline{h3} constructed before it, and invokes \lstinline{mul_eq_zero}, of type \lstinline{a * b = 0 ↔ a = 0 ∨ b = 0}, from \mathlib\ \cite{2025d}. The term \lstinline{h3} is of type \lstinline{a * b = 0}, for some \lstinline{a} and \lstinline{b}, and \lstinline{rw} updates \lstinline{h3} to be of type \lstinline{a = 0 ∨ b = 0}.

With the tactics understood, the no-cloning proof In \lean\ closely follows from the natural language proof. Below, we start by proving three smaller propositions --- \lstinline{h1}, \lstinline{h2}, and \lstinline{h3} --- which respectively correspond to (the first equality of) Eq.~\eqref{eq:no-cloning2}, Eq.~\eqref{eq:no-cloning3}, and Eq.~\eqref{eq:no-cloning4}. Propositions \lstinline{h1} and \lstinline{h2} are shown through external theorems, and \lstinline{h3} follows from \lstinline{h1} and \lstinline{h2} in \lstinline{exact congr(Subtype.val $h1).trans h2}. Afterwards, we use \lstinline{h3} and \lstinline{mul_eq_zero} to conclude that either $\braket{\psi}{\varphi} = 0$ or $\braket{\psi}{\varphi} = 1$.\footnote{A small disclaimer is that we work with pure density matrices in the code, such as \lstinline{pure ψ}, hence expressions involving \lstinline{trace.re}. In translating to the natural language proof, the change $\braket{\psi}{\phi} \to p = \abs{\braket{\psi}{\phi}}^2$ is implied.} Finally, the hypothesis \lstinline{H} is called to exclude the latter case, thus closing the goal with the term we started with, i.e. $\braket{\psi}{\varphi} = 0$.

\begin{codeListingBox}{No-cloning theorem}
\scriptsize\textit{QuantumInfo/Finite/Unitary.lean}
\begin{lstlisting}
theorem no_cloning {ψ φ f : Ket d} {U : Matrix.unitaryGroup n ℂ}
  (hψ : U ◃ pure (ψ ⊗ f) = pure (ψ ⊗ ψ))
  (hφ : U ◃ pure (φ ⊗ f) = pure (φ ⊗ φ))
  (H : ⟪pure ψ, pure φ⟫ < (1 : ℝ)) :
    ⟪pure ψ, pure φ⟫ = (0 : ℝ) := by
  set ρψ := pure ψ
  set ρφ := pure φ
  have h1 : ⟪ρψ, ρφ⟫ * ⟪ρψ, ρφ⟫ = ⟪pure (ψ ⊗ ψ), pure (φ ⊗ φ)⟫ := by
    grind only [pure_prod_pure, prod_inner_prod]
  have h2 : (⟪pure (ψ ⊗ ψ), pure (φ ⊗ φ)⟫ : ℝ) = ⟪U ◃ pure (ψ ⊗ f), U ◃ pure (φ ⊗ f)⟫ := by
    grind only [pure_prod_pure]
  replace h2 : ((pure (ψ ⊗ ψ)).m * (pure (φ ⊗ φ)).m).trace.re = (ρψ.m * ρφ.m).trace.re := by
    convert ← h2
    simp +zetaDelta only [inner_U_conj, pure_prod_pure, prod]
    simp [inner, HermitianMat.inner_eq_re_trace, ← mul_kronecker_mul, pure_mul_self, trace_kronecker]
  have h3 : ((ρψ.m * ρφ.m).trace.re) * ((ρψ.m * ρφ.m).trace.re - 1) = 0 := by
    rw [mul_sub, sub_eq_zero, mul_one]
    exact congr(Subtype.val $h1).trans h2
  rw [mul_eq_zero] at h3
  -- See Fig. (*@\ref{fig:infoview_no-cloning}@*) for the proof state here (line 99, column 24)
  apply h3.resolve_right
  exact sub_ne_zero_of_ne H.ne
\end{lstlisting}
\end{codeListingBox}

\begin{figure}
    \centering
    \includegraphics[width=0.73\linewidth]{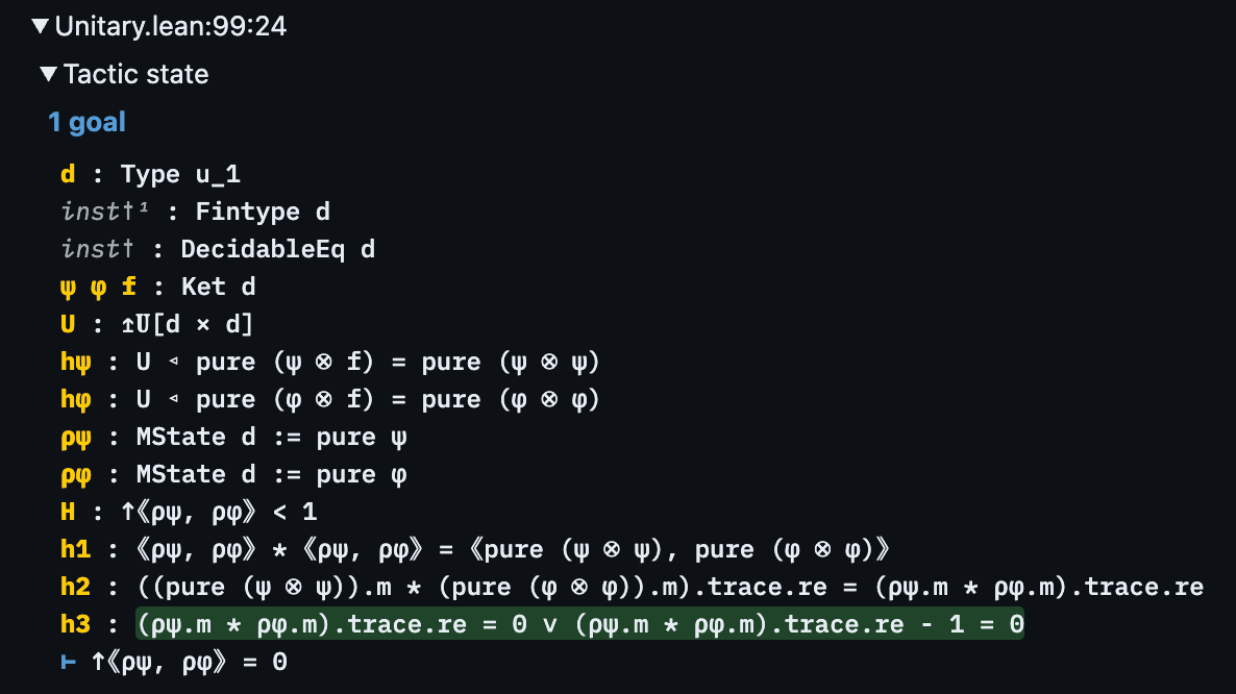}
    \caption{\lean's InfoView showing the state of the no-cloning proof just after proving proposition \lstinline{h3}, as indicated by the comment in the code listing. All terms from the local scope are explicitly stated along with their definitions, with the current goal ($\vdash \langle\psi, \varphi\rangle = 0$) at the end. As the proof evolves, both the list of hypotheses and the goal can change.}
    \label{fig:infoview_no-cloning}
\end{figure}


\subsection{Generalized quantum Stein's lemma}\label{sec:GQSL_explanation}





The Generalized Quantum Stein's Lemma relaxes the i.i.d.\ assumption going into the regular Quantum Stein's Lemma. The set $\mathcal{S}_n$ of allowed states comprising the alternative hypothesis now includes states other than many copies, $\sigma^{\otimes n}$ of $\sigma$.

The task of hypothesis testing requires, beyond these alternative states, the null hypothesis state $\rho$ from which we construct $\rho^{\otimes n}$. A successful test is the result of applying a two-outcome POVM $\{T_\rho, \openone - T_\rho\}$, where $0 \leq T_\rho \leq \openone$, $T_\rho \in \mathcal{B}(\mathcal{H}^{\otimes n})$, where $T_\rho$ signals the receipt of the state $\rho^{\otimes n}$, and $\openone - T_\rho$ signals ``not $\rho$.''

We closely follow the exposition of the protocol in Ref.~\cite{hayashi2024}. We chose to formalize this proof as opposed to the one in Ref.~\cite{lami2025}, because the latter requires lifting the problem to infinite-dimensional Hilbert spaces. These are technically more challenging and not as well developed in \mathlib\ when compared to finite-dimensional spaces. Any hypothesis testing protocol admits errors of type-I and type-II. An error of type-I occurs when the received quantum system was in state $\rho^{\otimes n} \in \mathcal{D}(\mathcal{H}^{\otimes n})$, but the test returned the outcome corresponding to $\openone - T_\rho$. This error occurs with probability $\alpha_n = \Tr[(\openone - T_\rho)\rho^{\otimes n}]$. The type-II error happens with probability $\beta_n = \max\{\Tr[T_\rho \sigma_n] \mid \sigma_n \in \mathcal{S}_n\}$, and corresponds to measuring the outcome of $T_\rho$, when the actual state was $\sigma_n$. In Ref.~\cite{hayashi2024} considers the worst possible scenario for the type-II error, hence the maximization over the alternative states.

Before making the precise statement of the GQSL below, we state its aim: by fixing an acceptable error probability $0 \leq \varepsilon \leq 1$ that bounds the type-I error, $\alpha_n \leq \varepsilon$, it is possible to bound the type-II error given an optimal POVM $T_\rho$. The exponential suppression of $\beta_n$, with increasing number of copies $n$, is shown to have as exponent the regularized relative entropy,
\begin{equation}
    R = \lim_{n\to \infty} \frac{1}{n} \min_{\sigma \in \mathcal{S}_n} D(\rho^{\otimes n}\|\sigma) ,
\end{equation}
where $\mathcal{S}_n$ is the subset of states that generalizes the i.i.d.\ copies of the original quantum Stein's lemma. These are sets with the following properties.
\begin{enumerate}
    \item $\mathcal{S}_n$ is a compact and convex set (hence closed for finite-dimensional $\mathcal{H}$).
    \item $\mathcal{S}_n$ is closed under tensor products, that is, $\rho_1 \otimes \rho_2 \in \mathcal{S}_{m+n}$ for $\rho_1 \in \mathcal{S}_m$ and $\rho_2 \in \mathcal{S}_n$.
    \item There exists a full-rank state belonging to the set, $\sigma_1 \in \mathcal{S}_1$.
\end{enumerate}
Although independent of quantum resource theory, these properties are naturally satisfied by the sets of free states of the $n$-party Hilbert space $\Hilb^{\otimes n}$ for many QRTs. 

Finally, we define the set of restricted POVMs,
\begin{equation}
    \mathcal{T}_{\varepsilon, \rho} = \{ T_\rho \in \mathcal{B}(\mathcal{H}^{\otimes n}) \mid
    0 \leq T_\rho \leq \openone, \Tr[(\openone - T_\rho)\rho^{\otimes n}] \leq \varepsilon\} ,
\end{equation}
%
and the associated $T_\rho$-minimal type-II error,
\begin{equation}
    \beta_{\varepsilon}(\rho \| \mathcal{S}) = \min_{T \in \mathcal{T}_{\varepsilon, \rho}} \max_{\sigma \in \mathcal{S}} \Tr[T \sigma] .
\end{equation}

These are the ingredients to introduce the generalized quantum Stein's lemma, in natural language.



\begin{theorem}[Generalized quantum Stein's lemma \cite{hayashi2024}]\label{thm:GQSL}
For any $\varepsilon\in(0,1)$ and any sequence $\{\mathcal{S}_n\}_n$ of sets of states satisfying Conditions 1, 2, and 3 above, we have
\begin{equation}
    \lim_{n \to \infty}- \frac{1}{n}\log \beta_\varepsilon(\rho^{\otimes n} \| \mathcal{S}_n) =
    \lim_{n\to \infty} \frac{1}{n} \min_{\sigma \in \mathcal{S}_n}D(\rho^{\otimes n}\|\sigma) .
    \label{ZXI_methods}
\end{equation}
\end{theorem}


\section{Outcomes of the Formalization}\label{sec:outcomes_of_the_formalization}
%

Our end goal is to formally verify all the statements made in the first half of Ref. \cite{hayashi2024}, leading up to the GQSL. In particular, we do not currently attempt to formalize the second half, which applies the GQSL to the second law of QRTs --- Theorem 2, ``Second law of QRTs for states.'' At present, statements of all the main definitions, lemmas and theorems of the paper are formalized in an almost one-to-one correspondence with their counterparts in Ref. \cite{hayashi2024}. In this sense, the {\em arguments of the paper} have been formally verified.

We have also built up an extensive body of underlying quantum theory, so that most of the theorems have an {\em end-to-end proof}. The remaining statements can be tracked by \lean's \lstinline{sorry} or \lstinline{axiom} command, and we have reduced it to a short list of extremely standard facts from quantum theory. Our remaining objective will be to proof these facts as well so the GQSL has an end-to-end proof, a project we expect to complete in a few coming months. The standards facts we need are:
\begin{itemize}
    \item The data processing inequality
    \item The additivity and lower semicontinuity of the relative entropy
    \item The continuity of sandwiched Rényi relative entropy $\tilde{D}_\alpha$ in $\alpha$
    \item The ``pinching Pythagoras'' theorem, that the pinching map gives a Pythagoras-like theorem for relative entropy
\end{itemize}
\lean\ facilitates the integration of unproved theorems into the development workflow through the keyword \lstinline{sorry}: stated theorems with proofs marked with \lstinline{sorry} are assumed to be true by the \lean\ compiler. Therefore, one can initially concentrate on stating the results, postponing their proofs.

To illustrate how the main result (Theorem \ref{thm:GQSL}) is formalized, we explain how it is written in the repository:

\begin{codeListingBox}{Generalized quantum Stein's lemma}
\scriptsize\textit{QuantumInfo/Finite/ResourceTheory/SteinsLemma.lean}
\begin{lstlisting}
theorem limit_hypotesting_eq_limit_rel_entropy (ρ : MState (H i)) (ε : Prob) 
(hε : 0 < ε ∧ ε < 1) :
∃ rate : ℝ≥0,
Filter.atTop.Tendsto (fun n ↦ —log β_ ε(ρ⊗^S[n]‖IsFree) / n) (calN rate)
∧
Filter.atTop.Tendsto (fun n ↦ (⨅ σ ∈ IsFree, bbD(ρ⊗^S[n]‖σ)) / n) (calN rate)
\end{lstlisting}
\end{codeListingBox}

The GQSL theorem is called \lstinline{limit_hypotesting_eq_limit_rel_entropy}, following \mathlib\ naming conventions \cite{mathlib_naming_conventions}. It takes three inputs, or assumptions: a mixed state $\rho \in \Hilb_i$, a probability $\varepsilon \in [0, 1]$, and a proof \lstinline{hε} that $\varepsilon \in (0, 1)$ (i.e. $\varepsilon$ is not zero nor one). More precisely, $\varepsilon$ is of type \lstinline{Prob}, which encapsulates a real number and a proof that it is an element of the interval $[0,1]$ in a single algebraic structure. 

After the colon, the theorem itself is stated. Since it is not possible to equate two limits in \mathlib\ per se\footnote{A limit does not always exist, and in non-Hausdorff topological spaces a limit may exist but not be unique. As such, it is poorly behaved as a ``function''. Instead, \mathlib\ works with predicates stating a sequence converges (\lstinline{Tendsto}) a particular value.}, we state it as the existence of a non-negative real number $\text{\lstinline{rate}} \in \R_{\geq 0}$ to which both limits converge. The first limit reads more naturally as
\begin{equation}
    \lim_{n \to \infty} -\log \beta_\varepsilon (\rho^{\otimes n} \Vert \mathcal{S}_n)/n = \text{\lstinline{rate}},
\end{equation}
and the second as
\begin{equation}
    \lim_{n \to \infty} \min_{\sigma \in \mathcal{S}_n} D(\rho^{\otimes n} \Vert \sigma)/n = \text{\lstinline{rate}}.
\end{equation}

Here, $\mathcal{S}_n$ is the set of free states of $\Hilb_i^{\otimes n}$ and is represented by \lstinline{IsFree} in the code; and $\beta_{\varepsilon}$ is a measure of discrimination between the null and alternate hypotheses, defined as the minimum type-II error $\beta_{\varepsilon}(\rho \Vert \mathcal{S}) \defeq \min_{T \in \mathcal{T}_{\varepsilon, \rho}} \max_{\sigma \in \mathcal{S}} \Tr[T \sigma]$ over POVMs $T$ with bounded type-I error $\Tr[(\one - T) \rho] \leq \varepsilon$. See Sec. \ref{sec:GQSL_explanation} for a more thorough explanation of these terms and the GQSL.

Along the way, we also proved many theorems applicable to a broader context than the GQSL's proof. Some of those were grouped into pull requests that were eventually added to the \mathlib\ repository  \cite{pr21351, pr21518, pr23960, pr24725, pr25194, pr27115, pr27118, pr29075, pr29228, pr29229, pr23960}.

\subsection{Imprecisions in the existing proof}\label{imprec}
Some technicalities exacted by type theory simply have no natural equivalent in written math. For instance, we needed to prove the fact that the relative entropy between two states is equal to the relative entropy between the {\em same} pair of states when interpreted under a different but equal Hilbert space\footnote{Proved in \lstinline{sandwichedRelRentropy\_heq\_congr}.}. These theorems --- and the fact that they need to be proved --- are, in our opinion, merely a byproduct of the embedding of quantum physics into type theoretic language. They are not substantially physically interesting.

Conversely, in any natural language proof, there are steps where more or less details could be provided. In the formalization process, we identified some places in \cite{hayashi2024} where a detail was not fully addressed.

Two issues have to deal with the handling of infinities that come from relative entropy. Relative entropy is typically given as a real number or $+\infty$, sometimes called the extended reals. Although it is easy to forget, the extended reals lack the nice algebraic properties of the reals, for instance $(a + b) - b \ne a$ in general, and indeed there is no agreed up on convention for what $\infty - \infty$ should evaluate to. In equation (S59) of \cite{hayashi2024}, two relative entropies are subtracted, with the implication that they can then be manipulated and cancelled from sides of an equation as reals numbers can be. This isn't true in general, and so of course \lean\ does not permit this manipulation\footnote{Unless of course the user proves that all numbers in play are finite, so that it can be reduced to a statement about the reals.}. We are able to carry out the algebra faithfully by instead keeping only sums, and avoiding subtracting infinities.

A later step of the proof involves applying Lemma 7. There are two quantities $R_1$ and $R_2$, which are each extended-real valued functions of a sequence of states, $\sigma_n$. Te goal is to show that there is a sequence $\sigma_n$ such the gap $R_2 - R_1$ becomes arbitrarily small. Lemma 7 says that, given a sequence $\sigma_n$, we can construct another sequence $\tilde{\sigma}_n$ with a gap at most $1 - \varepsilon$ times as large, for some $\varepsilon$ fixed by the setup; so by repeating this process we can squeeze the gap arbitrarily tight. They say that ``we start with any sequence'', but this forgets that we can have $R_1 < R_2 = +\infty$, since these again come from relative entropies. This means that we must first prove the existence of an initial sequence $\sigma_n$ with finite $R_2$, although this is easy in the context of the problem.

Other parts of \cite{hayashi2024} simply neglected to spell out every step. This doesn't constitute an error, but when details are omitted, a reader can be misled with an incorrect proof - an issue which is of course highly contextual and up to the reader. For instance, one step requires fact that there is a minimizer $\sigma$ that achieves $\inf_{\sigma \in S} D(\rho\| \sigma)$ when $S$ is compact. At first this might sound obvious since continuous functions are minimized on a compact set. But $D(\rho\| \cdot)$ is not continuous, only lower semicontinuous, a detail which a less informed reader could overlook. On the other hand, the justification given in the paper, ``the existence of a full-rank state in the sets''~\cite{hayashi2024v4}, is not necessary for the conclusion to hold. These are not errors, but in the formalization process we also uncover which hypotheses are {\em unnecessary} in some context. \lean\ gives a warning when a hypothesis is assumed or a fact is derived, and then never employed in the proof.

\section{Design and Development Choices}\label{sec:design_and_development_choices}
There were several design choices to be made during formalization: choices of mathematical convention, meaningful questions of what mathematical or physical scope to attack, and some difficult but pedestrian choices about good software engineering.

\subsection{Foundations of quantum theory}
There are several distinct foundations for ``quantum theory''. The most common is certainly that of bounded operators (equivalently, continuous linear maps) on Hilbert spaces, where {\em states} are the positive operator with unit trace. In the context of finite dimensions, the ``bounded" or ``continuous" prefixes are often dropped. A {\em Hilbert space with a basis} can be adopted as extra data, in which case there is a ``standard basis'' to refer to. Or one can forget the Hilbert space itself, leaving just a C$^*$-algebra\cite{ArvesonCStar}. By taking an enveloping W$^*$-algebra, one can instead move to von Neumann algebras\cite{BratteliVonNeumann,TakesakiVonNeumann} as a basis for quantum mechanics. Axiomatizations of quantum field theory like AQFT\cite{HaagAQFT} carry further data of spacetime regions associated to each local algebra. Still other foundations exist such as generalized probabilistic theories\cite{GPT-1,GPT-2,GPT-3}.

At the onset of the \leanqi\  repository, a medium-term goal was formalizing semantics from quantum computing. In the context of qubits and circuits, all computations are done in finite dimensions, and there is almost always a standard basis accessible as data. Basically every definition, such as a Pauli gate, stabilizer states, or the output distribution of a circuit, need to refer to a basis. This motivated the choice of {\em matrices} as the basic notion of a quantum mixed state, or observable.

Furthermore, we decided to view Hermitian matrices as their own first-class type. What does this mean? Physics notation is often heavily built around discouraging incorrect combination of data, what programming language theory calls {\em type safety}. For instance, it is obvious from appear that the expression
$$3\ket{\psi} - 4\bra{\phi}$$
is ill-typed, even though kets and bras are both ``just" vectors under the hood, and could be subtracted. This same functionality is supported in \lean, where our \lstinline{Bra} and \lstinline{Ket} types are both {\em coercible} to vectors, but {\em definitionally equal} to vectors. Similarly, if $H$ is a Hamiltonian and $U$ is a two-qubit unitary, an expression
$$H + 4U$$
should be alarming, even though these are both ``just" matrices. \lean's \mathlib\ already defines a special type \lstinline{Matrix.unitaryGroup} for type-safe manipulation of unitaries (e.g. multiplication is allowed, but not addition unless they are explicitly stripped to bare matrices first). Our repository defines, in a similar spirit, a \lstinline{HermitianMat} type for type-safe manipulation of Hermitian matrices, built upon the existing \mathlib\ predicate \lstinline{Matrix.IsHermitian}. This type permits addition and multiplication by reals, but not matrix multiplication; in place of multiplication, a typesafe \lstinline{conj} function is available for $H \mapsto A^\dagger H A$. Other benefits of the API are that the trace of the matrix is given as a manifestly real number, as opposed to a complex number which can subsequently proven real.

After then constructing the Loewner order, this leads to our definition of mixed state, perhaps the most central type in the repository:
\begin{codeListingBox}{Definition of a mixed state}
    \scriptsize\textit{QuantumInfo/Finite/MState.lean}
\begin{lstlisting}
structure MState (d : Type*) [Fintype d] [DecidableEq d] where
  M : HermitianMat d ℂ
  zero_le : 0 ≤ M
  tr : M.trace = 1
\end{lstlisting}
\end{codeListingBox}
This can be read as follows. An \lstinline{MState} is a data structure, which refers to another datatype \lstinline{d}; this other type \lstinline{d} must be finite and have a sensible notion of equality. (For instance, for a qubit, this other type could be a boolean, or the pair of strings ``Up'' and ``Down''; for a pair of qubits, it could be numbers from 1 or 4, or ordered pairs of booleans.) An \lstinline{MState} has a Hermitian matrix \lstinline{M} with rows and columns indexed by \lstinline{d}, and complex entries. The requirements are that \lstinline{M} be positive semidefinite and unit trace.

Why use a matrix instead of any of the other formulations? Simply because it is useful to carry around any accessible data when possible. By carrying around a standard basis labeled by \lstinline{d}, we give ourselves access to maximum contextual information, so that (for instance) the magic of a state is always accessible as a definition. We do lose access to the infinite dimensional quantum theory in the process, but this saves us a great many additional headaches in treating the correct topological subtleties, or ensuring that the quantities such as traces are always correctly convergent after each manipulation. Eventually formalizing an infinite dimensional theory will be necessary, but this will be best done with the ample learnings from a finite dimensional formalizations in hand.
    
    \subsection{Definition of quantum resource theory}

The generalized quantum Stein’s lemma is, first and foremost, a statement about quantum resource theories. There are several works that have given axiomatic descriptions of quantum resource theories~\cite{chitambar2019, fritzResources, spekkensResources}, but these do meaningfully differ, and all of them are essentially too imprecise to correspond directly to a \lean\ description. First, does a resource theory describe a set of free states, a set of free operations, or both (with some compatibility conditions?) Since we focus on the version of Stein’s lemma in terms of regularized entropies, the only actual data\footnote{In \lean, the distinction between data associated to a structure and proofs about the structure is central. For instance, a group is defined by the data of an identity, multiplication function, and inversion function; but one could imagine storing only the multiplication, with the existence of identity and inverse as propositions.} we need is the set of free states associated to each Hilbert space. We call this as a \lstinline{FreeStateTheory}, a structure we extend in \lstinline{ResourceTheory} to include a notion of free operation.

A resource theory also needs a {\em product} on Hilbert spaces. This is typically described as a tensor product, but this cannot be uniquely identifying. Consider the resource theory where Alice and Bob have distinct states, with LOCC operations as free. Then Alice having 2 qubits while Bob has 0 is a Hilbert space in the resource theory, and it is isomorphic to the Hilbert space where they each have 1 qubit. But when we take the product of two states with 1 qubit each, we need to know which Hilbert space we’re mapping to. Thus, the Hilbert spaces are {\em indexed} by some other type, and the product structure is a map which we require to be non-canonically isomorphic to the tensor product. This product cannot in general be associative, so we also require an associator.

If we permit a Hilbert space of dimension 1, then we have in fact reconstructed a monoidal category. We call resource theories with such a space {\em unital}. When we say that the resource theory becomes a (symmetric) monoidal category, note that this is a different sense than that of \cite{spekkensResources}, which also describes resource theories as a monoidal category, but their category has objects as quantum {\em states} as opposed to Hilbert {\em spaces}. This gives an operation-centric view of resource theories, while our \lstinline{FreeStateTheory} (which suffices for the proof) only mentions the allowed states.

In principle, every aspect of the proof in \cite{hayashi2024} appears to go through even in non-unital categories, and our initial efforts attempted to capture this. The proof development process was plagued by technical difficulties\footnote{Centrally: without a unit, a natural number power of states cannot be defined, as there is no zero power, and we only have positive integer powers. This meant poor interoperability with existing \mathlib\ code for natural numbers.}, though, and so we switched to unital (monoidal) resource theories for the remainder of the proof. When the proof is completed, extending it to non-unital resource theories is a goal for future work.

\subsection{Numerical convention}

Another central decision was whether to work with extended reals or not. We {\em do} use extended reals, as opposed to simply the reals, and we believe this improves the integrity of the proof at the expense of making some arguments involved. The impact of this choice requires understanding the notion of junk values, which requires some explanation. 

\subsubsection{Junk values}
A common convention in \lean\ is to adopt so-called ``junk values” when an output is otherwise ill-defined. For example, an infinite integral ``evaluates” to zero, as does the derivative of a non-differentiable function, or the limsup of an unbounded sequence. Division by zero also always produces zero. This may seem alarming, as this could lead to lead to theorems that don't mean what they appear to; but the threat is not as large as it may seem, since it only affects definitions used in the {\em statement} of a theorem. For example, the theorem (from core \lean) expressing that $n / n = 1$ for natural numbers $n$ reads as:

\begin{codeListingBox}{An example of a junk value condition}
\begin{lstlisting}
theorem Nat.div_self {n : Nat} (H : 0 < n) : n / n = 1 := by …
\end{lstlisting}
\end{codeListingBox}

This means that simplifying such an expression requires also proving that $n > 0$; if $n = 0$, then the junk value would imply that $0 / 0 = 0$. Does this risk ``compromising" any proof that involves division, since it could take advantage of this definition to mean something else? Suppose we had a \lean-verified proof of Fermat's Last Theorem\footnote{This theorem statement is taken from the ongoing FLT formalization project led by Kevin Buzzard, see \cite{BuzzardFLT}}:

\begin{codeListingBox}{A hypothetical proof of Fermat's Last Theorem}
\begin{lstlisting}
theorem PNat.pow_add_pow_ne_pow
  (x y z : ℕ+) (n : ℕ) (hn : n > 2) :
    x^n + y^n ≠ z^n := by …
\end{lstlisting}
\end{codeListingBox}

The {\em statement} of the theorem only requires the notions of addition, natural number powers, and non-equality. As such, it is insensitive to the existence of junk values - even though theorems concerning division, integration, etc. are certainly all part of the proof.

In earlier versions of \lean, junk values were avoided, and a division function would require a proof that the denominator is nonzero. This would lead to a function signature like $\mathbb{R} \to (y : \mathbb{R}) \to (y \ne 0) \to \mathbb{R}$. This guarantees that division is well-behaved wherever used, but leads to other issues involving {\em dependent rewrites}. For this reason, junk values are now generally preferred.

\subsubsection{Behavior of extended nonnegative reals}
The textbook definition of quantum relative entropy~\cite{nielsen2012} reads:
$$D(\rho || \sigma) \defeq \begin{cases} \operatorname{tr}(\rho(\log \rho - \log \sigma)) & \text{if } \operatorname{supp}(\rho) \subseteq \operatorname{supp}(\sigma) \\ \infty & \text{otherwise} \end{cases}$$
The {\em statement} of the (Generalized) Quantum Stein's Lemma does depend indirectly on the definition of relative entropy, so the way that we choose to formalize this definition can result in semantically distinct final theorem. That is, we {\em are} sensitive to junk values in this definition.

Following the convention of junk values in \mathlib, a reasonable way to define this would be
$$D(\rho || \sigma) \vcentcolon\stackrel{?}{=} \begin{cases} \operatorname{tr}(\rho(\log \rho - \log \sigma)) & \text{if } \operatorname{supp}(\rho) \subseteq \operatorname{supp}(\sigma) \\ 0 & \text{otherwise} \end{cases}$$
So that the value is always a finite\footnote{A matrix logarithm $\log \sigma$ could be expected to give infinite terms unless handled carefully, but \mathlib\ already defines $\log 0 = 0$, which guarantees a finite trace; and this agrees with the behavior we desire for relative entropies} real. This would have allowed us to work with the (very well supported) real numbers throughout the proof.

We opted to instead define the relative entropy as an {\em extended nonnegative real}, \lstinline{ENNReal} or $\mathbb{R}\ge0\infty$ in \lean, representing a number in the interval $[0,\infty]$ of the extended reals. This allows us to more accurately capture the semantics of the relative entropy, at the cost of ease of proof. For instance, the extended reals lack a continuous multiplication function:
$$\lim_{x\to \infty} \left( x * \frac{\pi}{x}\right) = \lim_{x \to \infty} \pi = \pi,\quad\textrm{but}$$
$$\left(\lim_{x\to \infty} x\right) * \left(\lim_{x\to \infty} \frac{\pi}{x}\right) = \infty * 0 = 0$$
This means that we cannot use the standard fact that $(\lim f)(\lim g) = \lim fg$ (when both limits exist) without separately dealing with the cases where $f$ or $g$ are either infinite or zero. The extended reals also do not form an additive group, as $(x + \infty) - \infty = \infty - \infty \ne x$ in general, making it harder to cancel equations.

If we had not adopted this version of the definition, we also could not have caught the error described in Section \ref{imprec}. By defining it properly with extended reals, we faced these larger mathematical difficulties in the manipulations of equations, and really captured the full physical semantics.

\subsection{Other mathematical hurdles}
Another more technical issue was finding several {\em definitional diamonds}, where one object inherits data from multiple sources. For us, this occurred with matrices inheriting a topology from multiple operator norms. The Hermitian matrix inner product naturally induces a norm, and a norm naturally induces a topology, but this topology is not {\em definitionally} the same as the one coming from the elementwise topology on matrices. Although the user can prove that these are the same, \lean\ doesn't recognize expressions involving one as also involving the other, and as a consequence extra care is required in the process of defining the inner product.

The Generalized Quantum Stein's Lemma involves several infima and suprema, for instance the optimal hypothesis testing rate is defined using an infimum over two-element POVMs. This is readily understood by humans to be equivalent to a infimum over a particular set of Hermitian matrices, or over an equivalent set of (not manifestly Hermitian) matrices; and the value of the infimum can be viewed as a probability, or a real number, or a complex number (since it is the inner product of two complex matrices). Each of these type changes requires proofs that the orderings are compatible and all sets are appropriately bounded (i.e. the infimum is finite), which considerably complicates the proof.

    \subsection{Comparison with other physics formalizations}
    Our work is not the first exploring formalizing physics in a formal theorem prover. \mathlib\ itself has a proof that the CHSH game has different values for commuting and noncommuting operators, phrased in the language of C$^*$-algebras. Some formalization of physics have been explored in Isabelle~\cite{Bordg2020,Stannett_2013}, but these have mostly been confined to a smaller scale. Two large projects stand out: the PhysLean~\cite{HepLean} library has developed a large collection of individual physics results, and there is a a verified implementation of Shor's algorithm in Rocq~\cite{PengShor}.

    There is a maxim in software engineering that a software {\em library} shouldn't be developed in isolation, but rather with an eye towards a particular application; for instance, the Rust programming language was heavily guided by a concurrent effort to develop a web browser in the system. Similarly, while \leanqi\ was initially a disconnected group of facts about quantum information, focusing on one theorem contributed to the formation of a coherent and integrated collection of theorems, with all the compatibility theorems and relations between predicates necessary to move between domains. This is similar to the focused approach adopted in the verified Shor's algorithm~\cite{PengShor}.

    But the most dramatic departure of our approach is working to verify a {\em new}, recent result, as opposed to standard textbook theorems.
    
    \section{Conclusion and outlook}\label{sec:conclusion_and_outlook}
    
    
    We have formally verified the proof of the Generalized Quantum Stein's Lemma as stated in Ref.~\cite{hayashi2024}. Exceptions to the steps leading to the GQSL are a handful of standard results in quantum information theory, such as the data processing inequality. Completing these will yield an end-to-end proof.
    
    This is, to our knowledge, the largest effort in formalizing a single theorem in physics using \lean. We anticipate that this and the accompanying code repository~\cite{meiburg2025} will foster productive collaboration between the \lean\ and formalization communities and the quantum information community.
    
    Next steps building up from the results shown include the removal of dependencies of the lemmas on other classic theorems, such as the data processing inequality, that have been stated as axioms. Proving these statements would greatly improve the reliability of a quantum information library written in \lean.
    
    More immediately applicable to the topic of GQSL, it would be also important to generalize the results to non-unital Quantum Resource Theories, and the results derived as corollaries of the GQSL, such as the Second Law of QRTs.
    

    
    This work establishes a formalized foundation for quantum physics and quantum information, and proves the usability of this foundation by proving one particular, highly non-trivial theorem. Different axiomatic constructions to quantum information theory can be considered in addition to the standard Hilbert space formulation~\cite{dirac1981,von2018mathematical,sakurai2020,nielsen2012}. Examples include generalized probabilistic theories~\cite{hardy_quantum_2001, barrett_information_2007, chiribella_probabilistic_2010, hardy_reconstructing_2013, muller_probabilistic_2021}, and C$^*$-algebraic or von Neumann algebraic theories~\cite{bratteli1987,bratteli1997}, the latter of which could be applied to algebraic quantum field theory~\cite{haag1996,fewster2019b}.
    All of these approaches can be simultaneously formalized, and benefit from being interconvertible and extendable. Furthermore, physics inspired approaches may also be considered, such as cases of Gaussian quantum mechanics, quantum optics, etc. The additional body of work on quantum information formalized in \lean\ means an efficient and new pathway for results and exchange of ideas, akin to community development in open source projects.
    

\begin{acknowledgments}
    We acknowledge contributions from Bolton Bailey (\textsf{@BoltonBailey}) and Lawrence Wu. We thank Alex May for organizing the workshop that kickstarted this project. We also thank Hayata Yamasaki and Masahito Hayashi for illuminating discussions and for writing the work on which our proof is based. R.R.S. thanks Ningping Cao for putting him in contact with A.M., and Mike and Ophelia Lazaridis for funding. L.A.L. acknowledges support from the Natural Sciences and Engineering Research Council of Canada (NSERC) under Discovery Grants No. RGPIN-2018-04380 and No. RGPIN-2020-04688. L.A.L. also acknowledges that this research was supported in part by grant NSF PHY-2309135 to the Kavli Institute for Theoretical Physics (KITP). A.M. acknowledges support under Discovery Grant No. RGPIN-2019-04198. This work was also supported by an Ontario Early Researcher Award. Research at Perimeter Institute is supported in part by the Government of Canada through the Department of Innovation, Science and Industry Canada and by the Province of Ontario through the Ministry of Colleges and Universities.
\end{acknowledgments}

\bibliography{bibliography}

\begin{thebibliography}{55}%
\makeatletter
\providecommand \@ifxundefined [1]{%
 \@ifx{#1\undefined}
}%
\providecommand \@ifnum [1]{%
 \ifnum #1\expandafter \@firstoftwo
 \else \expandafter \@secondoftwo
 \fi
}%
\providecommand \@ifx [1]{%
 \ifx #1\expandafter \@firstoftwo
 \else \expandafter \@secondoftwo
 \fi
}%
\providecommand \natexlab [1]{#1}%
\providecommand \enquote  [1]{``#1''}%
\providecommand \bibnamefont  [1]{#1}%
\providecommand \bibfnamefont [1]{#1}%
\providecommand \citenamefont [1]{#1}%
\providecommand \href@noop [0]{\@secondoftwo}%
\providecommand \href [0]{\begingroup \@sanitize@url \@href}%
\providecommand \@href[1]{\@@startlink{#1}\@@href}%
\providecommand \@@href[1]{\endgroup#1\@@endlink}%
\providecommand \@sanitize@url [0]{\catcode `\\12\catcode `\$12\catcode `\&12\catcode `\#12\catcode `\^12\catcode `\_12\catcode `\%12\relax}%
\providecommand \@@startlink[1]{}%
\providecommand \@@endlink[0]{}%
\providecommand \url  [0]{\begingroup\@sanitize@url \@url }%
\providecommand \@url [1]{\endgroup\@href {#1}{\urlprefix }}%
\providecommand \urlprefix  [0]{URL }%
\providecommand \Eprint [0]{\href }%
\providecommand \doibase [0]{https://doi.org/}%
\providecommand \selectlanguage [0]{\@gobble}%
\providecommand \bibinfo  [0]{\@secondoftwo}%
\providecommand \bibfield  [0]{\@secondoftwo}%
\providecommand \translation [1]{[#1]}%
\providecommand \BibitemOpen [0]{}%
\providecommand \bibitemStop [0]{}%
\providecommand \bibitemNoStop [0]{.\EOS\space}%
\providecommand \EOS [0]{\spacefactor3000\relax}%
\providecommand \BibitemShut  [1]{\csname bibitem#1\endcsname}%
\let\auto@bib@innerbib\@empty
\bibitem [{\citenamefont {Hiai}\ and\ \citenamefont {Petz}(1991)}]{hiai1991}%
  \BibitemOpen
  \bibfield  {author} {\bibinfo {author} {\bibfnamefont {F.}~\bibnamefont {Hiai}}\ and\ \bibinfo {author} {\bibfnamefont {D.}~\bibnamefont {Petz}},\ }\href {https://doi.org/10.1007/BF02100287} {\bibfield  {journal} {\bibinfo  {journal} {Commun.Math. Phys.}\ }\textbf {\bibinfo {volume} {143}},\ \bibinfo {pages} {99} (\bibinfo {year} {1991})}\BibitemShut {NoStop}%
\bibitem [{\citenamefont {Ogawa}\ and\ \citenamefont {Nagaoka}(2000)}]{ogawa2000}%
  \BibitemOpen
  \bibfield  {author} {\bibinfo {author} {\bibfnamefont {T.}~\bibnamefont {Ogawa}}\ and\ \bibinfo {author} {\bibfnamefont {H.}~\bibnamefont {Nagaoka}},\ }\href {https://doi.org/10.1109/18.887855} {\bibfield  {journal} {\bibinfo  {journal} {IEEE Trans. Inf. Theory}\ }\textbf {\bibinfo {volume} {46}},\ \bibinfo {pages} {2428} (\bibinfo {year} {2000})}\BibitemShut {NoStop}%
\bibitem [{\citenamefont {Brand{\~a}o}\ and\ \citenamefont {Plenio}(2010)}]{brandao2010}%
  \BibitemOpen
  \bibfield  {author} {\bibinfo {author} {\bibfnamefont {F.~G. S.~L.}\ \bibnamefont {Brand{\~a}o}}\ and\ \bibinfo {author} {\bibfnamefont {M.~B.}\ \bibnamefont {Plenio}},\ }\href {https://doi.org/10.1007/s00220-010-1005-z} {\bibfield  {journal} {\bibinfo  {journal} {Commun. Math. Phys.}\ }\textbf {\bibinfo {volume} {295}},\ \bibinfo {pages} {791} (\bibinfo {year} {2010})},\ \Eprint {https://arxiv.org/abs/0904.0281} {arXiv:0904.0281 [quant-ph]} \BibitemShut {NoStop}%
\bibitem [{\citenamefont {Chitambar}\ and\ \citenamefont {Gour}(2019)}]{chitambar2019}%
  \BibitemOpen
  \bibfield  {author} {\bibinfo {author} {\bibfnamefont {E.}~\bibnamefont {Chitambar}}\ and\ \bibinfo {author} {\bibfnamefont {G.}~\bibnamefont {Gour}},\ }\href {https://doi.org/10.1103/RevModPhys.91.025001} {\bibfield  {journal} {\bibinfo  {journal} {Rev. Mod. Phys.}\ }\textbf {\bibinfo {volume} {91}},\ \bibinfo {pages} {025001} (\bibinfo {year} {2019})},\ \Eprint {https://arxiv.org/abs/1806.06107} {arXiv:1806.06107 [quant-ph]} \BibitemShut {NoStop}%
\bibitem [{\citenamefont {Berta}\ \emph {et~al.}(2023)\citenamefont {Berta}, \citenamefont {Brand{\~a}o}, \citenamefont {Gour}, \citenamefont {Lami}, \citenamefont {Plenio}, \citenamefont {Regula},\ and\ \citenamefont {Tomamichel}}]{berta2023}%
  \BibitemOpen
  \bibfield  {author} {\bibinfo {author} {\bibfnamefont {M.}~\bibnamefont {Berta}}, \bibinfo {author} {\bibfnamefont {F.~G. S.~L.}\ \bibnamefont {Brand{\~a}o}}, \bibinfo {author} {\bibfnamefont {G.}~\bibnamefont {Gour}}, \bibinfo {author} {\bibfnamefont {L.}~\bibnamefont {Lami}}, \bibinfo {author} {\bibfnamefont {M.~B.}\ \bibnamefont {Plenio}}, \bibinfo {author} {\bibfnamefont {B.}~\bibnamefont {Regula}},\ and\ \bibinfo {author} {\bibfnamefont {M.}~\bibnamefont {Tomamichel}},\ }\href {https://doi.org/10.22331/q-2023-09-07-1103} {\bibfield  {journal} {\bibinfo  {journal} {Quantum}\ }\textbf {\bibinfo {volume} {7}},\ \bibinfo {pages} {1103} (\bibinfo {year} {2023})},\ \Eprint {https://arxiv.org/abs/2205.02813} {arXiv:2205.02813 [quant-ph]} \BibitemShut {NoStop}%
\bibitem [{\citenamefont {Hayashi}\ and\ \citenamefont {Yamasaki}(2024{\natexlab{a}})}]{hayashi2024}%
  \BibitemOpen
  \bibfield  {author} {\bibinfo {author} {\bibfnamefont {M.}~\bibnamefont {Hayashi}}\ and\ \bibinfo {author} {\bibfnamefont {H.}~\bibnamefont {Yamasaki}},\ }\href {https://doi.org/10.48550/arXiv.2408.02722} {\bibinfo {title} {Generalized {{Quantum Stein}}'s {{Lemma}} and {{Second Law}} of {{Quantum Resource Theories}}}} (\bibinfo {year} {2024}{\natexlab{a}}),\ \Eprint {https://arxiv.org/abs/2408.02722v3} {arXiv:2408.02722v3} \BibitemShut {NoStop}%
\bibitem [{\citenamefont {Lami}(2025)}]{lami2025}%
  \BibitemOpen
  \bibfield  {author} {\bibinfo {author} {\bibfnamefont {L.}~\bibnamefont {Lami}},\ }\href {https://doi.org/10.1109/TIT.2025.3543610} {\bibfield  {journal} {\bibinfo  {journal} {IEEE Trans. Inform. Theory}\ }\textbf {\bibinfo {volume} {71}},\ \bibinfo {pages} {4454} (\bibinfo {year} {2025})},\ \Eprint {https://arxiv.org/abs/2408.06410} {arXiv:2408.06410 [quant-ph]} \BibitemShut {NoStop}%
\bibitem [{\citenamefont {de~Moura}\ and\ \citenamefont {Ullrich}(2021)}]{moura2021}%
  \BibitemOpen
  \bibfield  {author} {\bibinfo {author} {\bibfnamefont {L.}~\bibnamefont {de~Moura}}\ and\ \bibinfo {author} {\bibfnamefont {S.}~\bibnamefont {Ullrich}},\ }in\ \href {https://doi.org/10.1007/978-3-030-79876-5_37} {\emph {\bibinfo {booktitle} {Autom. {{Deduc}}. -- {{CADE}} 28}}},\ \bibinfo {editor} {edited by\ \bibinfo {editor} {\bibfnamefont {A.}~\bibnamefont {Platzer}}\ and\ \bibinfo {editor} {\bibfnamefont {G.}~\bibnamefont {Sutcliffe}}}\ (\bibinfo  {publisher} {Springer International Publishing},\ \bibinfo {address} {Cham},\ \bibinfo {year} {2021})\ pp.\ \bibinfo {pages} {625--635}\BibitemShut {NoStop}%
\bibitem [{\citenamefont {{de Moura}}\ \emph {et~al.}(2015)\citenamefont {{de Moura}}, \citenamefont {Kong}, \citenamefont {Avigad}, \citenamefont {{van Doorn}},\ and\ \citenamefont {{von Raumer}}}]{demoura2015}%
  \BibitemOpen
  \bibfield  {author} {\bibinfo {author} {\bibfnamefont {L.}~\bibnamefont {{de Moura}}}, \bibinfo {author} {\bibfnamefont {S.}~\bibnamefont {Kong}}, \bibinfo {author} {\bibfnamefont {J.}~\bibnamefont {Avigad}}, \bibinfo {author} {\bibfnamefont {F.}~\bibnamefont {{van Doorn}}},\ and\ \bibinfo {author} {\bibfnamefont {J.}~\bibnamefont {{von Raumer}}},\ }in\ \href {https://doi.org/10.1007/978-3-319-21401-6_26} {\emph {\bibinfo {booktitle} {Autom. {{Deduc}}. - {{CADE-25}}}}},\ \bibinfo {editor} {edited by\ \bibinfo {editor} {\bibfnamefont {A.~P.}\ \bibnamefont {Felty}}\ and\ \bibinfo {editor} {\bibfnamefont {A.}~\bibnamefont {Middeldorp}}}\ (\bibinfo  {publisher} {Springer International Publishing},\ \bibinfo {address} {Cham},\ \bibinfo {year} {2015})\ pp.\ \bibinfo {pages} {378--388}\BibitemShut {NoStop}%
\bibitem [{202(2025{\natexlab{a}})}]{2025d}%
  \BibitemOpen
  \href {https://github.com/leanprover-community/mathlib4} {\bibinfo {title} {Leanprover-community/mathlib4}},\ \bibinfo {howpublished} {leanprover-community} (\bibinfo {year} {2025}{\natexlab{a}})\BibitemShut {NoStop}%
\bibitem [{202(2025{\natexlab{b}})}]{2025c}%
  \BibitemOpen
  \href {https://github.com/leanprover/lean4} {\bibinfo {title} {Leanprover/lean4}},\ \bibinfo {howpublished} {Lean} (\bibinfo {year} {2025}{\natexlab{b}})\BibitemShut {NoStop}%
\bibitem [{\citenamefont {{The Mathlib Community}}(2020)}]{themathlibcommunity2020}%
  \BibitemOpen
  \bibfield  {author} {\bibinfo {author} {\bibnamefont {{The Mathlib Community}}},\ }in\ \href {https://doi.org/10.1145/3372885.3373824} {\emph {\bibinfo {booktitle} {Proc. 9th {{ACM SIGPLAN Int}}. {{Conf}}. {{Certif}}. {{Programs Proofs}}}}}\ (\bibinfo  {publisher} {ACM},\ \bibinfo {address} {New Orleans LA USA},\ \bibinfo {year} {2020})\ pp.\ \bibinfo {pages} {367--381}\BibitemShut {NoStop}%
\bibitem [{\citenamefont {Meiburg}(2025{\natexlab{a}})}]{meiburg2025}%
  \BibitemOpen
  \bibfield  {author} {\bibinfo {author} {\bibfnamefont {A.}~\bibnamefont {Meiburg}},\ }\href {https://github.com/Timeroot/Lean-QuantumInfo} {\bibinfo {title} {Timeroot/{{Lean-QuantumInfo}}}} (\bibinfo {year} {2025}{\natexlab{a}})\BibitemShut {NoStop}%
\bibitem [{\citenamefont {Harrison}(2009)}]{harrison_handbook_2009}%
  \BibitemOpen
  \bibfield  {author} {\bibinfo {author} {\bibfnamefont {J.}~\bibnamefont {Harrison}},\ }\href@noop {} {\emph {\bibinfo {title} {Handbook of {{Practical Logic}} and {{Automated Reasoning}}}}}\ (\bibinfo  {publisher} {Cambridge University Press},\ \bibinfo {year} {2009})\BibitemShut {NoStop}%
\bibitem [{\citenamefont {{Lean Community}}()}]{lean_papers}%
  \BibitemOpen
  \bibfield  {author} {\bibinfo {author} {\bibnamefont {{Lean Community}}},\ }\href {https://leanprover-community.github.io/papers.html} {\bibinfo {title} {Formalization papers using lean}},\ \bibinfo {howpublished} {\url{https://leanprover-community.github.io/papers.html}}\BibitemShut {NoStop}%
\bibitem [{\citenamefont {Wootters}\ and\ \citenamefont {Zurek}(1982)}]{wootters1982}%
  \BibitemOpen
  \bibfield  {author} {\bibinfo {author} {\bibfnamefont {W.~K.}\ \bibnamefont {Wootters}}\ and\ \bibinfo {author} {\bibfnamefont {W.~H.}\ \bibnamefont {Zurek}},\ }\href {https://doi.org/10.1038/299802a0} {\bibfield  {journal} {\bibinfo  {journal} {Nature}\ }\textbf {\bibinfo {volume} {299}},\ \bibinfo {pages} {802} (\bibinfo {year} {1982})}\BibitemShut {NoStop}%
\bibitem [{\citenamefont {{Lean community}}()}]{mathlib_naming_conventions}%
  \BibitemOpen
  \bibfield  {author} {\bibinfo {author} {\bibnamefont {{Lean community}}},\ }\href {https://leanprover-community.github.io/contribute/naming.html} {\bibinfo {title} {Mathlib naming conventions}},\ \bibinfo {howpublished} {\url{https://leanprover-community.github.io/contribute/naming.html}}\BibitemShut {NoStop}%
\bibitem [{\citenamefont {Meiburg}(2025{\natexlab{b}})}]{pr21351}%
  \BibitemOpen
  \bibfield  {author} {\bibinfo {author} {\bibfnamefont {A.}~\bibnamefont {Meiburg}},\ }\href {https://github.com/leanprover-community/mathlib4/pull/21351} {\bibinfo {title} {{feat(Analysis/RCLike/Basic): PosMulReflectLE}}},\ \bibinfo {howpublished} {\url{https://github.com/leanprover-community/mathlib4/pull/21351}} (\bibinfo {year} {2025}{\natexlab{b}})\BibitemShut {NoStop}%
\bibitem [{\citenamefont {Meiburg}(2025{\natexlab{c}})}]{pr21518}%
  \BibitemOpen
  \bibfield  {author} {\bibinfo {author} {\bibfnamefont {A.}~\bibnamefont {Meiburg}},\ }\href {https://github.com/leanprover-community/mathlib4/pull/21518} {\bibinfo {title} {{feat(Logic/Equiv): Upgrade arrowProdEquivProdArrow to dependent types}}},\ \bibinfo {howpublished} {\url{https://github.com/leanprover-community/mathlib4/pull/21518}} (\bibinfo {year} {2025}{\natexlab{c}})\BibitemShut {NoStop}%
\bibitem [{\citenamefont {Meiburg}(2025{\natexlab{d}})}]{pr23960}%
  \BibitemOpen
  \bibfield  {author} {\bibinfo {author} {\bibfnamefont {A.}~\bibnamefont {Meiburg}},\ }\href {https://github.com/leanprover-community/mathlib4/pull/23960} {\bibinfo {title} {{feat(Analysis/RCLike): RCLike.StarModule R}}},\ \bibinfo {howpublished} {\url{https://github.com/leanprover-community/mathlib4/pull/23960}} (\bibinfo {year} {2025}{\natexlab{d}})\BibitemShut {NoStop}%
\bibitem [{\citenamefont {Meiburg}(2025{\natexlab{e}})}]{pr24725}%
  \BibitemOpen
  \bibfield  {author} {\bibinfo {author} {\bibfnamefont {A.}~\bibnamefont {Meiburg}},\ }\href {https://github.com/leanprover-community/mathlib4/pull/24725} {\bibinfo {title} {{feat(LinearAlgebra/Matrix/PosDef): Matrix.PosSemidef.det\_nonneg}}},\ \bibinfo {howpublished} {\url{https://github.com/leanprover-community/mathlib4/pull/24725}} (\bibinfo {year} {2025}{\natexlab{e}})\BibitemShut {NoStop}%
\bibitem [{\citenamefont {Meiburg}(2025{\natexlab{f}})}]{pr25194}%
  \BibitemOpen
  \bibfield  {author} {\bibinfo {author} {\bibfnamefont {A.}~\bibnamefont {Meiburg}},\ }\href {https://github.com/leanprover-community/mathlib4/pull/25194} {\bibinfo {title} {{chore(Analysis/SpecialFunctions/ContinuousFunctionalCalculus/ExpLog): weaken positivity hypothesis}}},\ \bibinfo {howpublished} {\url{https://github.com/leanprover-community/mathlib4/pull/25194}} (\bibinfo {year} {2025}{\natexlab{f}})\BibitemShut {NoStop}%
\bibitem [{\citenamefont {Meiburg}(2025{\natexlab{g}})}]{pr27115}%
  \BibitemOpen
  \bibfield  {author} {\bibinfo {author} {\bibfnamefont {A.}~\bibnamefont {Meiburg}},\ }\href {https://github.com/leanprover-community/mathlib4/pull/27115} {\bibinfo {title} {{feat(ENNReal/Lemmas): limsup/liminf of f+g when either f or g tends to zero}}},\ \bibinfo {howpublished} {\url{https://github.com/leanprover-community/mathlib4/pull/27115}} (\bibinfo {year} {2025}{\natexlab{g}})\BibitemShut {NoStop}%
\bibitem [{\citenamefont {Meiburg}(2025{\natexlab{h}})}]{pr27118}%
  \BibitemOpen
  \bibfield  {author} {\bibinfo {author} {\bibfnamefont {A.}~\bibnamefont {Meiburg}},\ }\href {https://github.com/leanprover-community/mathlib4/pull/27118} {\bibinfo {title} {{feat(Matrix/Charpoly/Eigs): Roots of Matrix.charpoly are the eigenvalues}}},\ \bibinfo {howpublished} {\url{https://github.com/leanprover-community/mathlib4/pull/27118}} (\bibinfo {year} {2025}{\natexlab{h}})\BibitemShut {NoStop}%
\bibitem [{\citenamefont {Meiburg}(2025{\natexlab{i}})}]{pr29075}%
  \BibitemOpen
  \bibfield  {author} {\bibinfo {author} {\bibfnamefont {A.}~\bibnamefont {Meiburg}},\ }\href {https://github.com/leanprover-community/mathlib4/pull/29075} {\bibinfo {title} {{feat(Analysis/Convex): Lifting convex sets along scalar towers}}},\ \bibinfo {howpublished} {\url{https://github.com/leanprover-community/mathlib4/pull/29075}} (\bibinfo {year} {2025}{\natexlab{i}})\BibitemShut {NoStop}%
\bibitem [{\citenamefont {Meiburg}(2025{\natexlab{j}})}]{pr29228}%
  \BibitemOpen
  \bibfield  {author} {\bibinfo {author} {\bibfnamefont {A.}~\bibnamefont {Meiburg}},\ }\href {https://github.com/leanprover-community/mathlib4/pull/29228} {\bibinfo {title} {{HEq iff Exists a cast}}},\ \bibinfo {howpublished} {\url{https://github.com/leanprover-community/mathlib4/pull/29228}} (\bibinfo {year} {2025}{\natexlab{j}})\BibitemShut {NoStop}%
\bibitem [{\citenamefont {Meiburg}(2025{\natexlab{k}})}]{pr29229}%
  \BibitemOpen
  \bibfield  {author} {\bibinfo {author} {\bibfnamefont {A.}~\bibnamefont {Meiburg}},\ }\href {https://github.com/leanprover-community/mathlib4/pull/29229} {\bibinfo {title} {{feat(Logic/Equiv/Defs): Equiv.trans\_cancel\_left / right}}},\ \bibinfo {howpublished} {\url{https://github.com/leanprover-community/mathlib4/pull/29229}} (\bibinfo {year} {2025}{\natexlab{k}})\BibitemShut {NoStop}%
\bibitem [{\citenamefont {Hayashi}\ and\ \citenamefont {Yamasaki}(2024{\natexlab{b}})}]{hayashi2024v4}%
  \BibitemOpen
  \bibfield  {author} {\bibinfo {author} {\bibfnamefont {M.}~\bibnamefont {Hayashi}}\ and\ \bibinfo {author} {\bibfnamefont {H.}~\bibnamefont {Yamasaki}},\ }\href {https://doi.org/10.48550/arXiv.2408.02722} {\bibinfo {title} {Generalized {{Quantum Stein}}'s {{Lemma}} and {{Second Law}} of {{Quantum Resource Theories}}}} (\bibinfo {year} {2024}{\natexlab{b}}),\ \Eprint {https://arxiv.org/abs/2408.02722} {arXiv:2408.02722 [quant-ph]} \BibitemShut {NoStop}%
\bibitem [{\citenamefont {Arveson}(1998)}]{ArvesonCStar}%
  \BibitemOpen
  \bibfield  {author} {\bibinfo {author} {\bibfnamefont {W.}~\bibnamefont {Arveson}},\ }\href@noop {} {\emph {\bibinfo {title} {An Invitation to {C*-Algebras}}}},\ \bibinfo {edition} {1st}\ ed.,\ Graduate Texts in Mathematics\ (\bibinfo  {publisher} {Springer},\ \bibinfo {address} {New York, NY},\ \bibinfo {year} {1998})\BibitemShut {NoStop}%
\bibitem [{\citenamefont {Bratteli}\ and\ \citenamefont {Robinson}(1981)}]{BratteliVonNeumann}%
  \BibitemOpen
  \bibfield  {author} {\bibinfo {author} {\bibfnamefont {O.}~\bibnamefont {Bratteli}}\ and\ \bibinfo {author} {\bibfnamefont {D.~W.}\ \bibnamefont {Robinson}},\ }\href@noop {} {\emph {\bibinfo {title} {Operator algebras and quantum statistical mechanics {II}}}},\ Theoretical and Mathematical Physics\ (\bibinfo  {publisher} {Springer},\ \bibinfo {address} {Berlin, Germany},\ \bibinfo {year} {1981})\BibitemShut {NoStop}%
\bibitem [{\citenamefont {Takesaki}(2002)}]{TakesakiVonNeumann}%
  \BibitemOpen
  \bibfield  {author} {\bibinfo {author} {\bibfnamefont {M.}~\bibnamefont {Takesaki}},\ }\href@noop {} {\emph {\bibinfo {title} {Theory of operator algebras {III}}}},\ \bibinfo {edition} {2003rd}\ ed.,\ Encyclopaedia of Mathematical Sciences\ (\bibinfo  {publisher} {Springer},\ \bibinfo {address} {Berlin, Germany},\ \bibinfo {year} {2002})\BibitemShut {NoStop}%
\bibitem [{\citenamefont {Haag}(1996{\natexlab{a}})}]{HaagAQFT}%
  \BibitemOpen
  \bibfield  {author} {\bibinfo {author} {\bibfnamefont {R.}~\bibnamefont {Haag}},\ }\href@noop {} {\emph {\bibinfo {title} {Local quantum physics}}},\ \bibinfo {edition} {2nd}\ ed.,\ Theoretical and Mathematical Physics\ (\bibinfo  {publisher} {Springer},\ \bibinfo {address} {Berlin, Germany},\ \bibinfo {year} {1996})\BibitemShut {NoStop}%
\bibitem [{\citenamefont {Barrett}(2007{\natexlab{a}})}]{GPT-1}%
  \BibitemOpen
  \bibfield  {author} {\bibinfo {author} {\bibfnamefont {J.}~\bibnamefont {Barrett}},\ }\href@noop {} {\bibfield  {journal} {\bibinfo  {journal} {Phys. Rev. A}\ }\textbf {\bibinfo {volume} {75}} (\bibinfo {year} {2007}{\natexlab{a}})}\BibitemShut {NoStop}%
\bibitem [{\citenamefont {Scandolo}\ \emph {et~al.}(2021)\citenamefont {Scandolo}, \citenamefont {Salazar}, \citenamefont {Korbicz},\ and\ \citenamefont {Horodecki}}]{GPT-2}%
  \BibitemOpen
  \bibfield  {author} {\bibinfo {author} {\bibfnamefont {C.~M.}\ \bibnamefont {Scandolo}}, \bibinfo {author} {\bibfnamefont {R.}~\bibnamefont {Salazar}}, \bibinfo {author} {\bibfnamefont {J.~K.}\ \bibnamefont {Korbicz}},\ and\ \bibinfo {author} {\bibfnamefont {P.}~\bibnamefont {Horodecki}},\ }\href@noop {} {\bibfield  {journal} {\bibinfo  {journal} {Phys. Rev. Res.}\ }\textbf {\bibinfo {volume} {3}} (\bibinfo {year} {2021})}\BibitemShut {NoStop}%
\bibitem [{\citenamefont {Spekkens}(2007)}]{GPT-3}%
  \BibitemOpen
  \bibfield  {author} {\bibinfo {author} {\bibfnamefont {R.~W.}\ \bibnamefont {Spekkens}},\ }\href@noop {} {\bibfield  {journal} {\bibinfo  {journal} {Phys. Rev. A}\ }\textbf {\bibinfo {volume} {75}} (\bibinfo {year} {2007})}\BibitemShut {NoStop}%
\bibitem [{\citenamefont {Fritz}(2017)}]{fritzResources}%
  \BibitemOpen
  \bibfield  {author} {\bibinfo {author} {\bibfnamefont {T.}~\bibnamefont {Fritz}},\ }\href {https://doi.org/10.1017/S0960129515000444} {\bibfield  {journal} {\bibinfo  {journal} {Mathematical Structures in Computer Science}\ }\textbf {\bibinfo {volume} {27}},\ \bibinfo {pages} {850–938} (\bibinfo {year} {2017})}\BibitemShut {NoStop}%
\bibitem [{\citenamefont {Coecke}\ \emph {et~al.}(2016)\citenamefont {Coecke}, \citenamefont {Fritz},\ and\ \citenamefont {Spekkens}}]{spekkensResources}%
  \BibitemOpen
  \bibfield  {author} {\bibinfo {author} {\bibfnamefont {B.}~\bibnamefont {Coecke}}, \bibinfo {author} {\bibfnamefont {T.}~\bibnamefont {Fritz}},\ and\ \bibinfo {author} {\bibfnamefont {R.~W.}\ \bibnamefont {Spekkens}},\ }\href {https://doi.org/https://doi.org/10.1016/j.ic.2016.02.008} {\bibfield  {journal} {\bibinfo  {journal} {Information and Computation}\ }\textbf {\bibinfo {volume} {250}},\ \bibinfo {pages} {59} (\bibinfo {year} {2016})},\ \bibinfo {note} {quantum Physics and Logic}\BibitemShut {NoStop}%
\bibitem [{\citenamefont {Buzzard}\ and\ \citenamefont {Taylor}()}]{BuzzardFLT}%
  \BibitemOpen
  \bibfield  {author} {\bibinfo {author} {\bibfnamefont {K.}~\bibnamefont {Buzzard}}\ and\ \bibinfo {author} {\bibfnamefont {R.}~\bibnamefont {Taylor}},\ }\href {https://imperialcollegelondon.github.io/FLT/blueprint.pdf} {\bibinfo {title} {Towards a lean proof of fermat’s last theorem}}\BibitemShut {NoStop}%
\bibitem [{\citenamefont {Nielsen}\ and\ \citenamefont {Chuang}(2012)}]{nielsen2012}%
  \BibitemOpen
  \bibfield  {author} {\bibinfo {author} {\bibfnamefont {M.~A.}\ \bibnamefont {Nielsen}}\ and\ \bibinfo {author} {\bibfnamefont {I.~L.}\ \bibnamefont {Chuang}},\ }\href {https://doi.org/10.1017/CBO9780511976667} {\emph {\bibinfo {title} {Quantum {{Computation}} and {{Quantum Information}}: 10th {{Anniversary Edition}}}}},\ \bibinfo {edition} {1st}\ ed.\ (\bibinfo  {publisher} {Cambridge University Press},\ \bibinfo {address} {Cambridge},\ \bibinfo {year} {2012})\BibitemShut {NoStop}%
\bibitem [{\citenamefont {Bordg}\ \emph {et~al.}(2020)\citenamefont {Bordg}, \citenamefont {Lachnitt},\ and\ \citenamefont {He}}]{Bordg2020}%
  \BibitemOpen
  \bibfield  {author} {\bibinfo {author} {\bibfnamefont {A.}~\bibnamefont {Bordg}}, \bibinfo {author} {\bibfnamefont {H.}~\bibnamefont {Lachnitt}},\ and\ \bibinfo {author} {\bibfnamefont {Y.}~\bibnamefont {He}},\ }\href {https://doi.org/10.1007/s10817-020-09584-7} {\bibfield  {journal} {\bibinfo  {journal} {Journal of Automated Reasoning}\ }\textbf {\bibinfo {volume} {65}},\ \bibinfo {pages} {691–709} (\bibinfo {year} {2020})}\BibitemShut {NoStop}%
\bibitem [{\citenamefont {Stannett}\ and\ \citenamefont {Németi}(2013)}]{Stannett_2013}%
  \BibitemOpen
  \bibfield  {author} {\bibinfo {author} {\bibfnamefont {M.}~\bibnamefont {Stannett}}\ and\ \bibinfo {author} {\bibfnamefont {I.}~\bibnamefont {Németi}},\ }\href {https://doi.org/10.1007/s10817-013-9292-7} {\bibfield  {journal} {\bibinfo  {journal} {Journal of Automated Reasoning}\ }\textbf {\bibinfo {volume} {52}},\ \bibinfo {pages} {361–378} (\bibinfo {year} {2013})}\BibitemShut {NoStop}%
\bibitem [{\citenamefont {Tooby-Smith}(2024)}]{HepLean}%
  \BibitemOpen
  \bibfield  {author} {\bibinfo {author} {\bibfnamefont {J.}~\bibnamefont {Tooby-Smith}},\ }\href {https://doi.org/10.1016/j.cpc.2024.109457} {\bibfield  {journal} {\bibinfo  {journal} {Comput. Phys. Commun.}\ ,\ \bibinfo {pages} {109457}} (\bibinfo {year} {2024})}\BibitemShut {NoStop}%
\bibitem [{\citenamefont {Peng}\ \emph {et~al.}(2023)\citenamefont {Peng}, \citenamefont {Hietala}, \citenamefont {Tao}, \citenamefont {Li}, \citenamefont {Rand}, \citenamefont {Hicks},\ and\ \citenamefont {Wu}}]{PengShor}%
  \BibitemOpen
  \bibfield  {author} {\bibinfo {author} {\bibfnamefont {Y.}~\bibnamefont {Peng}}, \bibinfo {author} {\bibfnamefont {K.}~\bibnamefont {Hietala}}, \bibinfo {author} {\bibfnamefont {R.}~\bibnamefont {Tao}}, \bibinfo {author} {\bibfnamefont {L.}~\bibnamefont {Li}}, \bibinfo {author} {\bibfnamefont {R.}~\bibnamefont {Rand}}, \bibinfo {author} {\bibfnamefont {M.}~\bibnamefont {Hicks}},\ and\ \bibinfo {author} {\bibfnamefont {X.}~\bibnamefont {Wu}},\ }\bibfield  {journal} {\bibinfo  {journal} {Proceedings of the National Academy of Sciences}\ }\textbf {\bibinfo {volume} {120}},\ \href {https://doi.org/10.1073/pnas.2218775120} {10.1073/pnas.2218775120} (\bibinfo {year} {2023})\BibitemShut {NoStop}%
\bibitem [{\citenamefont {Dirac}(1981)}]{dirac1981}%
  \BibitemOpen
  \bibfield  {author} {\bibinfo {author} {\bibfnamefont {P.~A.~M.}\ \bibnamefont {Dirac}},\ }\href@noop {} {\emph {\bibinfo {title} {The {{Principles}} of {{Quantum Mechanics}}}}}\ (\bibinfo  {publisher} {Clarendon Press},\ \bibinfo {year} {1981})\BibitemShut {NoStop}%
\bibitem [{\citenamefont {{von Neumann}}\ \emph {et~al.}(2018)\citenamefont {{von Neumann}}, \citenamefont {Beyer},\ and\ \citenamefont {Wheeler}}]{von2018mathematical}%
  \BibitemOpen
  \bibfield  {author} {\bibinfo {author} {\bibfnamefont {J.}~\bibnamefont {{von Neumann}}}, \bibinfo {author} {\bibfnamefont {R.}~\bibnamefont {Beyer}},\ and\ \bibinfo {author} {\bibfnamefont {N.}~\bibnamefont {Wheeler}},\ }\href {https://books.google.ca/books?id=z3g9DwAAQBAJ} {\emph {\bibinfo {title} {Mathematical Foundations of Quantum Mechanics: {{New}} Edition}}},\ Princeton Landmarks in Mathematics and Physics\ (\bibinfo  {publisher} {Princeton University Press},\ \bibinfo {year} {2018})\BibitemShut {NoStop}%
\bibitem [{\citenamefont {Sakurai}\ and\ \citenamefont {Napolitano}(2020)}]{sakurai2020}%
  \BibitemOpen
  \bibfield  {author} {\bibinfo {author} {\bibfnamefont {J.~J.}\ \bibnamefont {Sakurai}}\ and\ \bibinfo {author} {\bibfnamefont {J.}~\bibnamefont {Napolitano}},\ }\href {https://doi.org/10.1017/9781108587280} {\emph {\bibinfo {title} {Modern {{Quantum Mechanics}}}}},\ \bibinfo {edition} {3rd}\ ed.\ (\bibinfo  {publisher} {Cambridge University Press},\ \bibinfo {year} {2020})\BibitemShut {NoStop}%
\bibitem [{\citenamefont {Hardy}(2001)}]{hardy_quantum_2001}%
  \BibitemOpen
  \bibfield  {author} {\bibinfo {author} {\bibfnamefont {L.}~\bibnamefont {Hardy}},\ }\href {https://doi.org/10.48550/arXiv.quant-ph/0101012} {\bibinfo {title} {Quantum {{Theory From Five Reasonable Axioms}}}} (\bibinfo {year} {2001}),\ \Eprint {https://arxiv.org/abs/quant-ph/0101012} {arXiv:quant-ph/0101012} \BibitemShut {NoStop}%
\bibitem [{\citenamefont {Barrett}(2007{\natexlab{b}})}]{barrett_information_2007}%
  \BibitemOpen
  \bibfield  {author} {\bibinfo {author} {\bibfnamefont {J.}~\bibnamefont {Barrett}},\ }\href {https://doi.org/10.1103/PhysRevA.75.032304} {\bibfield  {journal} {\bibinfo  {journal} {Physical Review A}\ }\textbf {\bibinfo {volume} {75}},\ \bibinfo {pages} {032304} (\bibinfo {year} {2007}{\natexlab{b}})}\BibitemShut {NoStop}%
\bibitem [{\citenamefont {Chiribella}\ \emph {et~al.}(2010)\citenamefont {Chiribella}, \citenamefont {D'Ariano},\ and\ \citenamefont {Perinotti}}]{chiribella_probabilistic_2010}%
  \BibitemOpen
  \bibfield  {author} {\bibinfo {author} {\bibfnamefont {G.}~\bibnamefont {Chiribella}}, \bibinfo {author} {\bibfnamefont {G.~M.}\ \bibnamefont {D'Ariano}},\ and\ \bibinfo {author} {\bibfnamefont {P.}~\bibnamefont {Perinotti}},\ }\href {https://doi.org/10.1103/PhysRevA.81.062348} {\bibfield  {journal} {\bibinfo  {journal} {Physical Review A}\ }\textbf {\bibinfo {volume} {81}},\ \bibinfo {pages} {062348} (\bibinfo {year} {2010})}\BibitemShut {NoStop}%
\bibitem [{\citenamefont {Hardy}(2013)}]{hardy_reconstructing_2013}%
  \BibitemOpen
  \bibfield  {author} {\bibinfo {author} {\bibfnamefont {L.}~\bibnamefont {Hardy}},\ }\href {https://doi.org/10.48550/arXiv.1303.1538} {\bibinfo {title} {Reconstructing quantum theory}} (\bibinfo {year} {2013}),\ \Eprint {https://arxiv.org/abs/1303.1538} {arXiv:1303.1538 [quant-ph]} \BibitemShut {NoStop}%
\bibitem [{\citenamefont {M{\"u}ller}(2021)}]{muller_probabilistic_2021}%
  \BibitemOpen
  \bibfield  {author} {\bibinfo {author} {\bibfnamefont {M.}~\bibnamefont {M{\"u}ller}},\ }\href {https://doi.org/10.21468/SciPostPhysLectNotes.28} {\bibfield  {journal} {\bibinfo  {journal} {SciPost Physics Lecture Notes}\ ,\ \bibinfo {pages} {028}} (\bibinfo {year} {2021})}\BibitemShut {NoStop}%
\bibitem [{\citenamefont {Bratteli}\ and\ \citenamefont {Robinson}(1987)}]{bratteli1987}%
  \BibitemOpen
  \bibfield  {author} {\bibinfo {author} {\bibfnamefont {O.}~\bibnamefont {Bratteli}}\ and\ \bibinfo {author} {\bibfnamefont {D.~W.}\ \bibnamefont {Robinson}},\ }\href {https://doi.org/10.1007/978-3-662-02520-8} {\emph {\bibinfo {title} {Operator {{Algebras}} and {{Quantum Statistical Mechanics}} 1}}}\ (\bibinfo  {publisher} {Springer Berlin Heidelberg},\ \bibinfo {address} {Berlin, Heidelberg},\ \bibinfo {year} {1987})\BibitemShut {NoStop}%
\bibitem [{\citenamefont {Bratteli}\ and\ \citenamefont {Robinson}(1997)}]{bratteli1997}%
  \BibitemOpen
  \bibfield  {author} {\bibinfo {author} {\bibfnamefont {O.}~\bibnamefont {Bratteli}}\ and\ \bibinfo {author} {\bibfnamefont {D.~W.}\ \bibnamefont {Robinson}},\ }\href {https://doi.org/10.1007/978-3-662-03444-6} {\emph {\bibinfo {title} {Operator {{Algebras}} and {{Quantum Statistical Mechanics}} 2}}}\ (\bibinfo  {publisher} {Springer Berlin Heidelberg},\ \bibinfo {address} {Berlin, Heidelberg},\ \bibinfo {year} {1997})\BibitemShut {NoStop}%
\bibitem [{\citenamefont {Haag}(1996{\natexlab{b}})}]{haag1996}%
  \BibitemOpen
  \bibfield  {author} {\bibinfo {author} {\bibfnamefont {R.}~\bibnamefont {Haag}},\ }\href {https://doi.org/10.1007/978-3-642-61458-3} {\emph {\bibinfo {title} {Local {{Quantum Physics}}}}}\ (\bibinfo  {publisher} {Springer Berlin Heidelberg},\ \bibinfo {address} {Berlin, Heidelberg},\ \bibinfo {year} {1996})\BibitemShut {NoStop}%
\bibitem [{\citenamefont {Fewster}\ and\ \citenamefont {Rejzner}(2019)}]{fewster2019b}%
  \BibitemOpen
  \bibfield  {author} {\bibinfo {author} {\bibfnamefont {C.~J.}\ \bibnamefont {Fewster}}\ and\ \bibinfo {author} {\bibfnamefont {K.}~\bibnamefont {Rejzner}},\ }\href {https://arxiv.org/abs/1904.04051} {\bibinfo {title} {Algebraic quantum field theory -- an introduction}} (\bibinfo {year} {2019}),\ \Eprint {https://arxiv.org/abs/1904.04051} {arXiv:1904.04051 [hep-th]} \BibitemShut {NoStop}%
\end{thebibliography}%

\appendix

\section{\texorpdfstring{Other examples of \lean\ proofs}{Other examples of Lean proofs}}
\label{app:other_examples}

Here, we provide three other examples of theorems proved in the \leanqi\ repository. They are all elementary in nature, and serve to illustrate how proofs are written in \lean.

\subsection{Hilbert-Schmidt inner product of positive semidefinite matrices}

The first example showcases a basic fact of the Hilbert-Schmidt inner product $\langle A, B \rangle = \Tr[A^\dagger B]$. It is a complex inner product when defined for complex matrices $A$ and $B$, but becomes a real inner product when restricted to the real subspace of Hermitian matrices. In this subspace, the conjugate transpose can be dropped and we have $\langle A, B \rangle = \Tr[A B]$.

Our goal is to prove that the Hilbert-Schmidt inner product between two positive semidefinite matrices $A$ and $B$ is nonnegative, that is
\begin{equation}
    0 \leq A \text{ and } 0 \leq B \implies 0 \leq \langle A, B\rangle \defeq \Tr[A B]. 
\end{equation}
In natural language, the proof is simply
\begin{align}
    \langle A, B \rangle & = \Tr[A B] \label{eq:inner_prod1}\\
    & = \Tr[\sqrt{A} B \sqrt{A}] \label{eq:inner_prod2} \\
    & = \Tr[\sqrt{A}^\dagger B \sqrt{A}] \label{eq:inner_prod3} \\
    & \geq 0
\end{align}
At the last step, we use the fact that conjugating a positive matrix $B$ by another matrix $\sqrt{A}$ gives another positive matrix, and that the trace of a positive matrix is nonnegative.

The code listing below roughly follows the proof above. There, the inner product between $A$ and $B$ is written as \lstinline{A.inner B}. The type of $A$ and $B$ is defined earlier to be \lstinline{HermitianMat n t}, so they are already Hermitian matrices. 

In the third line of the proof, the first rewrite (\lstinline{rw}) expression \lstinline{inner_eq_re_trace} replaces the goal \lstinline{0 ≤ A.inner B} with \lstinline{0 ≤ RCLike.re (↑A * ↑B).trace}, which translates to $0 \leq \text{Re} \Tr[A B]$. The real part is necessary for the expected types to match: the inner product between Hermitian matrices should be real, but the trace of complex matrices is, in general, a complex number. Indeed, the Hermitian matrices $A$ and $B$ are cast to complex matrices with \lstinline{↑}. This detail, however, does not change the structure of the rest of the proof, and we will omit the ``real part'' operator in what follows.

In the rest of the rewrite operation in the third line of the proof, we convert $\Tr[A B]$ into $\Tr[\sqrt{A} B \sqrt{A}]$ by invoking that $A = \sqrt{A} \sqrt{A}$ (\lstinline{←hA.sqrt_mul_self}) and the cyclicity of the trace (\lstinline{Matrix.trace_mul_cycle}), thus arriving at Eq.~\eqref{eq:inner_prod2}. In the fourth line, we use \lstinline{nth_rewrite 1} replace the first appearance of $\sqrt{A}$ with $\sqrt{A}^\dagger$; and, finally, we use that $B \geq 0 \Rightarrow \sqrt{A}^\dagger B \sqrt{A} \geq 0$ (\lstinline{hB.conjTranspose_mul_mul_same}) and that $\sqrt{A}^\dagger B \sqrt{A} \geq 0 \Rightarrow \Tr[\sqrt{A}^\dagger B \sqrt{A}]$ (\lstinline{(...).trace_nonneg}). All of the statements employed above are prove

\begin{codeListingBox}{Hilbert-Schmidt inner product of positive semidefinite matrices is nonnegative}
\scriptsize\textit{QuantumInfo/ForMathlib/HermitianMat/Inner.lean}
\begin{lstlisting}
theorem inner_ge_zero (hA : 0 ≤ A) (hB : 0 ≤ B) : 0 ≤ A.inner B := by
  rw [zero_le_iff] at hA hB
  open Classical in
  rw [inner_eq_re_trace, ←hA.sqrt_mul_self, Matrix.trace_mul_cycle, Matrix.trace_mul_cycle]
  nth_rewrite 1 [←hA.posSemidef_sqrt.left]
  exact (RCLike.nonneg_iff.mp (hB.conjTranspose_mul_mul_same _).trace_nonneg).left
\end{lstlisting}
\end{codeListingBox}

\subsection{Trace distance between mixed states}
In this example, we examine the trace distance between mixed states $\rho$ and $\sigma$, defined as $D(\rho, \sigma) \defeq \frac{1}{2} \norm{\rho - \sigma}_1$, where $\norm{A}_1 \defeq \Tr \sqrt{A^\dagger A}$ is the trace norm. In particular, we prove that it is upper bounded by one:
\begin{equation}
    D(\rho, \sigma) \leq 1 
\end{equation}
In natural language, the proof is a simple application of the triangle inequality:
\begin{align}
    D(\rho, \sigma) & = \frac{1}{2} \norm{\rho - \sigma}_1 \\
    & \leq \frac{1}{2} (\norm{\rho}_1 + \norm{\sigma}_1) \\
    & = \frac{1}{2} (1 + 1) \\
    & = 1
\end{align}

The code listing below follows the proof above line-by-line. It also showcases the use of the \lstinline{calc} tactic. When the goal is to prove a transitive relation between two quantities --- in this case, that $D(\rho, \sigma) \leq 1$ --- \lstinline{calc} facilitates a step-by-step reasoning, wherein intermediate relations are proven separately.
\begin{codeListingBox}{The trace distance between two mixed states is at most 1}
\scriptsize\textit{QuantumInfo/Finite/Distance/TraceDistance.lean}
\begin{lstlisting}
theorem le_one : TrDistance ρ σ ≤ 1 :=
  calc TrDistance ρ σ -- A7, lhs
    _ = (1/2:ℝ) * (ρ.m - σ.m).traceNorm := by rfl -- A7, rhs
    _ ≤ (1/2:ℝ) * (ρ.m.traceNorm + σ.m.traceNorm) := by -- A8 and the proof of this inequality
      linarith [Matrix.traceNorm_triangleIneq' ρ.m σ.m]
    _ = (1/2:ℝ) * (1 + 1) := by -- A9 and its proof
      rw [ρ.traceNorm_eq_1, σ.traceNorm_eq_1]
    _ = 1 := by norm_num -- A10
\end{lstlisting}
\end{codeListingBox}

\subsection{Custom tactic for verifying quantum circuits}

A very important part of \lean's popularity is its highly extensible tactic system. A tactic is a command that automates part of a proof. The following excerpt shows our definition of a custom \lstinline{matrix_expand} tactic designed for checking quantum circuit equivalence by direct evaluation. This tactic is used in our repository to verify many simple facts, such as $HXH=Z$, in an automated way. In the excerpt below, it is used to verify two facts about controlled gates. In the first example, it resolves the entire proof on its own, and is able to reason about the ``abstract'' gate \lstinline{g₁}. In the second example, the tactic becomes part of a larger proof, showcasing the composability of \lean's tactic system.
\begin{codeListingBox}{A custom tactic for verifying quantum circuits}
\scriptsize\textit{QuantumInfo/Finite/Qubit/Basic.lean}
\begin{lstlisting}
/--
Proves goals equating small matrices by expanding out products and simplifying standard Real arithmetic.
-/
syntax (name := matrix_expand) "matrix_expand"
  (" [" ((simpStar <|> simpErase <|> simpLemma),*,?) "]")?
  (" with " rcasesPat+)? : tactic

macro_rules
  | `(tactic| matrix_expand $[[$rules,*]]? $[with $withArg*]?) => do
    let id1 := (withArg.getD ⟨[]⟩).getD 0 (← `(rcasesPat| _))
    let id2 := (withArg.getD ⟨[]⟩).getD 1 (← `(rcasesPat| _))
    let rules' := rules.getD ⟨#[]⟩
    `(tactic| (
      ext i j
      repeat rcases (i : Prod _ _) with ⟨i, $id1⟩
      repeat rcases (j : Prod _ _) with ⟨j, $id2⟩
      fin_cases i
      <;> fin_cases j
      <;> simp [Complex.ext_iff,
        Matrix.mul_apply, Fintype.sum_prod_type, Matrix.one_apply, field,
        $rules',* ]
      <;> norm_num
      <;> try field_simp
      <;> try ring_nf
      ))

/-- A controlled gate g₁ followed by controlled g₂ is the same as their controlled composition. -/
theorem controllize_mul (g₁ g₂ : U[k]) : C[g₁] * C[g₂] = C[g₁ * g₂] := by
  matrix_expand

/-- A controlled gate g, conjugated by X on the control qubit, is equivalent to applying g followed by a controlled g⁻¹. -/
theorem X_controllize_X : (X ⊗ 1) * C[g] * (X ⊗ 1) = (1 ⊗ g) * C[g⁻¹] := by
  matrix_expand [X, -Complex.ext_iff] with ki kj
  suffices (1 : Matrix k k ℂ) ki kj = (g * g⁻¹) ki kj by
    convert this
  simp
\end{lstlisting}
\end{codeListingBox}

\end{document}